%% file: main.tex
\documentclass{article}

\usepackage{arxiv}

\usepackage[utf8]{inputenc} % allow utf-8 input
\usepackage[T1]{fontenc}    % use 8-bit T1 fonts
\usepackage{hyperref}       % hyperlinks
\usepackage{url}            % simple URL typesetting
\usepackage{booktabs}       % professional-quality tables
\usepackage{amsfonts}       % blackboard math symbols
\usepackage{nicefrac}       % compact symbols for 1/2, etc.
\usepackage{microtype}      % microtypography
\usepackage{lipsum}

\usepackage{cite}
\usepackage{amsmath,amssymb,amsfonts}
\usepackage{algorithmic}
\usepackage{graphicx}
\usepackage{textcomp}
\usepackage{caption} 

\title{A Survey of Automatic Generation of Source Code Comments: Algorithms and Techniques}

\author{
  Xiaotao Song \\
  School of Software\\
  Taiyuan University of Technology, China\\
  \texttt{songxiaotao@tyut.edu.cn} \\
   \And
 Hailong Sun\thanks{Corresponding author}, Xu Wang and Jiafei Yan \\
  SKLSDE Lab, School of Computer Science and Engineering\\
  Beijing Advanced Innovation Center for Big Data and Brain Computing\\
  Beihang University, Beijing, China \\
  \texttt{sunhl@buaa.edu.cn} \\
}

\begin{document}
\maketitle

\begin{abstract}
%\lipsum[1]
As an integral part of source code files, code comments help improve program readability and comprehension. However, developers sometimes do not comment on their program code adequately due to the incurred extra efforts, lack of relevant knowledge, unawareness of the importance of code commenting or some other factors. 
As a result, code comments can be inadequate, absent or even mismatched with source code, which affects the understanding, reusing and the maintenance of software. To solve these problems of code comments, researchers have been concerned with generating code comments automatically. In this work, we aim at conducting a survey of automatic code commenting researches. First, we generally analyze the challenges and research framework of automatic generation of program comments. Second, we present the classification of representative algorithms, the design principles, strengths and weaknesses of each category of algorithms. %, the applicable condition and measures of further improving performance of algorithms. 
Meanwhile, %since researches on comment quality assessment play an important role in code commenting, 
we also provide an overview of the quality assessment of the generated comments. Finally, we summarize some future directions for advancing the techniques of automatic generation of code comments and the quality assessment of comments. 
\end{abstract}

% keywords can be removed

\input{introduction}
\input{overview}
\input{algorithm}
\input{quality}
\input{future}
\input{conclusion}

\bibliographystyle{unsrt} 
\bibliography{reference}



\end{document}

%% file: introduction.tex
\section{Introduction}
%\shl{The logic can be like this: (1) what are code comments, and why they are important or what roles they play in software engineering; (2) how to generate code comments in general; (3) what are the research challenges in automatic code generation, and the general state-of-the-art in this field; (4) the research objective of this work. (5) the organization of this paper. }

%\shl{Note: must pay attention to the logical connection among sentences!!}

Code comments, also called program annotations, are human-readable explanations or annotations of the source code of a computer program \cite{b1}, which mainly describe the functions and intentions of source code. Good comments can improve the readability of programs \cite{tenny1988program,tenny1985procedures,woodfield1981effect}, thus helping people comprehend programs. For instance, an early study \cite{tenny1985procedures} shows that comments can improve the readability of the banker algorithm used in operating systems. As a result, it has been widely acknowledged that comments play an important role in software development and maintenance   \cite{aggarwal2002integrated,tenny1988program,takang1996effects,woodfield1981effect,roehm2012professional}.
%Hence, along this path, 
%comments can also enhance the maintainability of programs \cite{aggarwal2002integrated}.
%Many studies \cite{tenny1988program,takang1996effects,woodfield1981effect,roehm2012professional} show that comments play an important role in comprehending programs.
% a program chosen to testify the effect of comments on the program readability.
%During the large amount of practice of software engineering, there exist some phenomena: on the one hand, 

%In software engineering practice, it is generally admitted by developers that accurate comments are the necessary companion of high quality code, and good comments will improve readability and maintainability of code \cite{aggarwal2002integrated}. 
However, writing high quality comments in practice during development is laborious and time-consuming for developers \cite{haouari2011good,kajko2005survey}. To deal with this issue, many efforts \cite{wong2013autocomment,wong2015clocom,hu2018deep,sridhara2010towards,rahman2015recommending} have been made towards automatically generating code comments.  
%\cite{wong2013autocomment,wong2015clocom,hu2018deep,sridhara2010towards} or partly automatically  \cite{rahman2015recommending}.
At the same time, researchers propose other approaches to improving the readability of programs too. For example, 
some researchers have tried to define identifiers with a long descriptive name in order to implement self-commented code \cite{fowler2018refactoring,allamanis2015suggesting}. However, it makes code comprehension more difficult \cite{binkley2008impact,liblit2006cognitive}. In general, automatic code commenting has become an important and challenging research direction in software engineering area.

%\shl{Now you start discussing how to generate code comments. it is OK.} 

Studies on code comments and readability of programs can be traced back to the 1980s \cite{tenny1988program,tenny1985procedures,woodfield1981effect} while the history of autocommenting just started in the last decade. % Generating code comments by computer had been a dream of developers for several decades. 
Existing methods are mainly based on machine learning or information retrieval techniques to generate comments for programs. %\shl{not true! a lot of work on IR}. 
The generation framework of code commenting is mainly composed of three parts. The first one is data preparation which prepares data for the commenting system. The second part is the representation of source code, which aims at capturing the structure and semantics of source code, such as information of structure, lexis, grammar, semantics, contexts, invocation relation and data dependency of source code. The third part is text generation, which is responsible for generating natural language sentences based on the information extracted from source code. %The task of text generation is constructing the information extracted from source code into natural language sentences. Concerning with the details that how each parts work together interactively will be discussed in section 2.

%During the process of automatic code comment generation, the topmost challenge is to design efficient algorithms of automatic commenting. During the practice of seeking for efficient algorithms of automatic comment generation for several decades, 
Along with the study of code comment researchers have found that the assessment of the quality of code comments is another important research problem, as the quality of generated comments % also called analytical comments, which produced by one except author of code,
is an important indicator for evaluating whether a commenting algorithm is efficient and effective.  %As mentioned above, inaccurate, unmatched, and wrong comments can not be beneficial for comprehending programs. Some comments may even bring bugs for software, which lower the quality of code.  
Thus designing appropriate criteria for code comment quality assessment is another challenge faced by automatic comment generation \cite{b5,khamis2010automatic,papineni2002bleu:,banerjee2005meteor:,denkowski2014meteor}. The task of assessment for quality of code comments involves the comparison and verification of various algorithms. 

%\xu{the structure of this paragraph is a little confused. First ... on the other hand...Secondly...}\shl{once again, you switch back to the topic of the quality issue. I agree with Xu there lacks a clear logic.}

%\shl{Now we should describe the state-of-the-art of automatic comment generation.}
%{In general, in the field of software engineering, existing studies on automatic code comment and the related areas mainly focus on algorithms of automatic comment generation. Researchers mostly adopted techniques from machine learning to automatically comment source code.}

There have been a lot of research efforts on automatic code commenting, especially from the overlapping community of software engineering and artificial intelligence. As a result, many papers have been published in top software engineering and artificial intelligence venues including IEEE ICSE, IEEE FSE, IEEE/ACM ASE, IEEE TSE, ACM TOSEM, EMSE, AAAI, IJCAI and so on.   
	%In recent years, software engineering international conferences, such as ICSE, ASE, FSE, MSR, and AI conferences, such as AAAI, IJCAI, and TSE, TOSEM, EMSE, etc., international top journals and conferences on data mining, have published a large number of papers about automatic code comment generation. %\shl{I think we can generally say there are a lot of research efforts from both software engineering community and ai community.} These papers present and share the achievements and results in their studies of automatic code comment generation. 
To our best knowledge, there are few efforts on the survey of studies in this field. In \cite{yang2019survey}, Yang et al. summarize the work on code comments from four aspects: code comment generation, classification of comments, the consistency of code and comments, and quality assessment of code comments. %The paper does not analyze the key problems of automatic code comment generation, and fails to give the whole picture of the development in this area. 
However, the paper does not discuss the principle of each algorithm in detail and fails to analyze the future research direction. Nazar et al. \cite{nazar2016summarizing} survey the studies on summarizing software artifacts, which include bug reports, mailing list, source code and developer discussions, where the task of summarizing source code is similar to code commenting because code summary can be viewed as a special type of comments. %, and summarization of source code is included in our paper. 
%The reason is that the summaries of source code can be viewed as a type of comments. %However, their work reviews extensively the summarization of various kinds of software artifacts, not exclusively summaries of source code. %\shl{any other surveys?}%\shl{I remember there are some survey papers? but they are far from thoroughly and comprehensively.}
In light of this, we aim at giving a comprehensive survey of the work on automatic code commenting for the following objectives: (1) giving researchers access to a catalogue of representative algorithms for automatic comment generation and providing new researchers with a good understanding of the state-of-the-art algorithms of automatic code commenting; (2) summarizing the main challenges and limitations of existing studies.  %\shl{(1) and (3) are the same.}% \xu{This paragraph is much important, which should describe the contributions more clearly}}

%\shl{Then we should describe the research objective of this work. Basically, there is almost no existing work on surveying this research topic although it has become a trending topic in software engineering. It is also necessary to discuss how the survey is done. For example, which conferences/journals are considered, how do you collect the papers.}
In this work, we present the motivation of automatic code comment generation first, analyze the main challenges, and describe the workflow of code commenting automatically. Next we discuss the three mainstream categories of algorithms of automatic comment generation, and show the potential trends in automatic comment techniques. This paper also summarizes the work on quality assessment of comments and presents the future direction accordingly. 
Note that in our survey we also investigate the work on generating code summaries, a short brief description of the code that is often viewed as a special type of comments. 

The rest of the paper is organized as follows. Section \ref{sec:overview} provides the motivation of automatic comment generation, and discusses the technical challenges. Section \ref{sec:algorithm} discusses the core ideas for comment generation techniques, and gives a summarization of all kinds of techniques. In Section \ref{sec:quality}, we discuss the problem of quality assessments of comments, with datasets used in different studies as our focus, and summarize the quality assessment criteria of code comments. The future directions on the automatic code comment generation will be discussed in Section \ref{sec:future}. In Section \ref{sec:conclusion}, we conclude the paper.

%\shl{One more question: what is the relationship of code comments and code summary? We must make it clear in this section.}

%% file: overview.tex
\section{Overview of automatic generation of code comments}
\label{sec:overview}
\subsection{Problem Statement}
Automatic code comment generation concerns the production of some textual descriptions of source code. % with the help of a computer program. Automatic code comment generation refers to choose the appropriate position automatically, and add descriptive natural language texts, named comments, for code automatically, aiming at describing the program logic and explaining the design intention of programs. 
The essential task is to translate the code written in programming languages into textual comments written in natural languages. Meanwhile, comments may describe not only the functions, but also the design intents of developers behind source code. 
In brief, automatic code commenting is to generate textual description written in natural languages automatically for source code by means of source code analysis, which can reveal the design intents, program logic, functionality of programs and the meanings of the related parameters, etc.

\subsection{CHALLENGES OF AUTOMATIC CODE COMMENTING AND RESEARCH FRAMEWORK}

Although the processes of different code commenting algorithms are not completely the same, the fundamental workflows are roughly similar, as shown in Figure 1.

%\Figure[t!](topskip=2pt, botskip=3pt, midskip=0pt){Figure1.pdf}
\begin{figure*}[t!]
	\centering
	\includegraphics[width=0.9\textwidth]{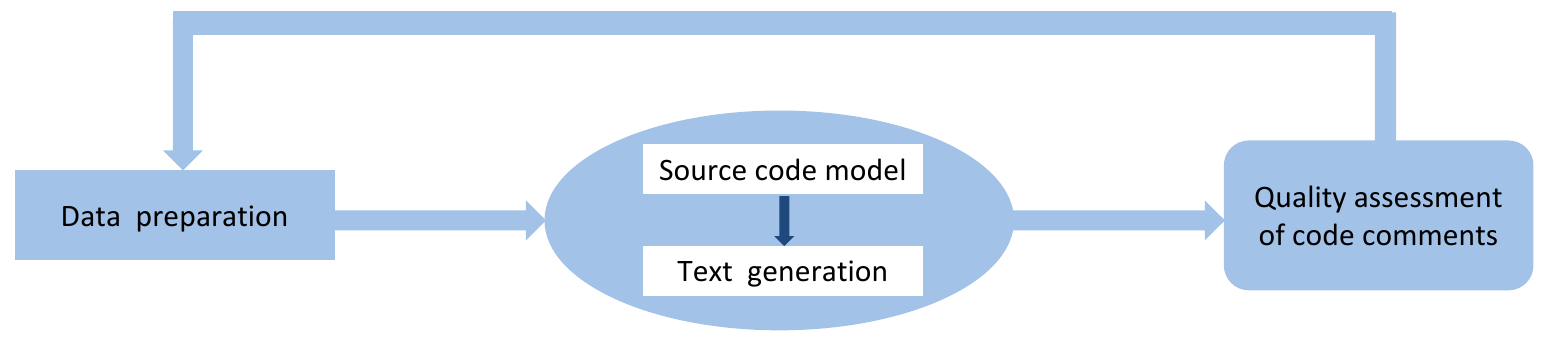}
    \caption{The general process of the automatic generation of code comments.}
		\label{fig1}
\end{figure*}
%\xu{all the figures seem to be unclear. They should use bigger pictures in the pdf format.}
%\includegraphics[scale=0.6]
%
The processing of automatic code commenting is usually performed in three steps. First, data collection to construct datasets for comment generation systems. These data are used for training, validating and testing models, extracting code and the corresponding comments, or extracting particular rules needed by a comment generation system. In order to collect these data, researchers often crawl or download them from open source communities or websites, e.g. Stack Overflow. Accordingly, the specific tasks in this step vary from algorithm to algorithm \cite{wong2013autocomment,wong2015clocom,chatterjee2017extracting} to some extent. For example, it is necessary for deep neural network based comment generation system \cite{hu2018deep,zheng2017code,allamanis2016convolutional,mou2016convolutional,hill2010integrating,iyer2016summarizing,wan2018improving} to build high quality datasets (i.e. source files) which contain code and the corresponding comments, so as to provide data for training, testing and verification of commenting algorithms.% If commenting systems adopt other machine learning algorithms to generate comments, extracting code segments and their descriptions\cite{chatterjee2017extracting}, and building high quality database which compromising of code and comment pairs are also necessary in this step.} 

%After data prepared, it comes into the key step of comments generation process, that is producing high quality comments for target source code in this step.
Second, comment generation through certain algorithms. This step can be divided into two subtasks, i.e. representation of source code and text generation. It involves varying processes depending on different algorithms of automatic code comment generation, which will be described in Section \ref{sec:algorithm}. %Finally, we will acquire the product of automatic commenting system, that is program comments.} 

%No matter which kind of algorithm adopted in automatic comment system, we all need to validate the performance of algorithms, and to evaluate the quality of generated comments. As a result, researches on assessing the quality of comments are the very important issues.
Third, the assessment of the generated comments in terms of their quality. Designing  practical and objective quality assessment criteria of comments directly affects the comparative results for different algorithms in  performance and quality assessment. There are two popular evaluation methods including human assessment and automatic assessment,  %The latter borrows assessment criteria and tools from the area of natural language processing. 
which will be discussed in Section \ref{sec:quality}. %\shl{Also need to explain the continuous improvement of comment quality after the assessment step.}

After assessment of code comments, commenting systems will take different actions depending on the assessment results of comments. If the amount and quality of comments generated by the commenting system is satisfactory, the process of commenting will stop. Otherwise, the commenting system will go back to the first step: preparing more and suitable data, and/or adjusting source code models, generating text and assessing the quality of code comments again, repeating this process till the need of code commenting is met.

\subsubsection{Challenges of Automatic Code Commenting}
As for automatic code commenting, the first thing is to build the source code model to express the structural, lexical, grammatical, semantic and context features of source code. Then, source code model is processed to yield the natural language comments. The third step is to evaluate the generated comments. However, generating satisfying code comments remains a challenging issue. The fundamental reason lies in the fact that programming languages are different from natural languages in nature. The difference between code and comments is two-fold: source code contains a large amount of information about classes, methods and parameters of methods, and at the same time has many nested structures and complex invocation relations; meanwhile, comments written in natural languages are unstructured, and expressed freely in form \cite{pinker1995language}. % and complex information they transferring\cite{hindle2012on}. %During the process of automatic program annotation, we should not only recognize accurately and express appropriately the information from source code, but also should solve the conversion of different representation forms  from source code to natural language comments. 
Consequently, automatic code commenting faces the following two challenges.

 \textbf{Challenge 1: automatic code commenting algorithms.}

At present, there exist many kinds of algorithms for automatic or semi-automatic code commenting. %\shl{program annotation is another way of describing code comment?} 
We summarize them into three main classes: information retrieval based algorithms, deep neural networks based algorithms and other automatic code comment generation algorithms (see Section \ref{sec:algorithm} for details).
	%\shl{where are they??})Most of algorithms aim at generating explanations comments or summary comments for source code, and several algorithms analyze the discussion posts on the Question and Answer websites or the communication letters from developers to generate comments about source code defects, concerns and restrictions etc.. However, no existing commenting algorithm could generate source code comments to describe design intentions, domain knowledge etc., or generate comments to meet the needs of accuracy and quantity. In order to generate comments which can serve all kinds of developers and maintenancers genuinely, it is necessary to improve the automatic code comment algorithms.}

%{There are two key steps in the process of automatic code commenting, one is building source code representation(or model), and the other is text generation.}
The two main issues in automatic commenting include source code representation and comment text generation.

  \textbf{Source code models.}
	{Among automatic comment generation studies, source code model is one of the core problems. 
	There are a number of different source code models including Abstract Syntax Trees (ASTs), parse trees, token contexts, Control Flow Graphs (CFGs), data flow, etc. %They are all able to represent source code from different perspectives.
	The source code models that have been used in autocommenting can be classified into three categories \cite{allamanis2018survey}. First, the token-based source code models, which mainly extract tokens such as key words and topics from source code, e.g.  %SLAMC (short for semantic language model for source code)
	the models in \cite{nguyen2013statistical,haiduc2010use}. In token-based models, source code is viewed as plain text, thus is often modeled as a bag of code tokens (BoT), or characters or bag of words (BoW). Information retrieval (IR) based commenting algorithms mainly adopt these models to represent source code. These models simply represent the lexical information of source code. Second, the syntax-based source code models, which model source code at level of abstract syntax trees (ASTs) \cite{b8}. These kinds of models are often used in deep neural network based commenting algorithms. Third, the other source code models, which represent code in forms that are fit for follow-up process. The models used in \cite{hill2010integrating,sridhara2010towards} belong to this category, such as Software Word Usage Model (SWUM). However, a combined model that can represent various information such as lexis, syntax and structure of source code is still missing though it has received some attention \cite{tufano2018deep}. As a result, seeking a comprehensive and effective model is an open research topic for source code commenting.}   %which is an instance of software bag of words(BOW)\cite{}, Resource Description Framework(RDF)\cite{rastkar2011generating}, which was proposed to describe resources in specific system for the first time. Second, semantic based source code models, which extracting semantic information from context enviroment of source code, such as Software Word Usage Model(SWUM)\cite{hill2010integrating} and the AST-based Neural Networks models\cite{hu2018deep,zheng2017code,allamanis2016convolutional,mou2016convolutional,hu2018summarizing}.}% In the early studies on code commenting, only the lexical information of identifiers and comments in source code were considered, and researchers represented source code as software bag of words(BOW)\cite{haiduc2010use} or corpus containing words and phrases, that is treating source code files as plain text, sets of words, with no sequence order, for instance SLAMC (short for semantic language model for source code)\cite{nguyen2013statistical}, an instance of the concept of BOW. Later, some researchers expressed the semantic information and invoke relations with Resource Description Framework(RDF)\cite{rastkar2011generating}, which was proposed to describe resources in specific system for the first time. And some studies represented the structural information, invoke relations, data dependency and words and phrases from source code as a combined model which can contain multiple levels, such as words level, phrase level and program structural level, and each level provides one kind of information in source code, such models as Software Word Usage Model(SWUM)\cite{hill2010integrating}. In recent two years, many researchers tried to exploit the deep neural network\cite{hu2018deep,zheng2017code,allamanis2016convolutional,mou2016convolutional,hu2018summarizing} to represent information from source code, which includes the lexicon information, invoke relation and the mapping relationships between code and comments sequences. These models mostly are based on ASTs, and are generally called AST-based Neural Networks models. } %\{b8}shl{Is there work on using AST to model the source code?}
	
	%{Generally, RDF model is fit for representing invoking relations between software entities. To some extent, SWUM model is a comprehensive model and well represents information from source code at the same time, eg words, data dependency, revoking relation, yet it is not fit for expressing the corresponding relation between code sequences and its comments. Through a large number of training, deep neural networks can also represent many kinds of information from source code, and its main advantage is to express the corresponding relation on sequences. However, there exits various kinds of information in source code, and the approach to recognizing each kind of information is different. Moreover, the expression forms suitable for varing information from source code are not same, and there exist complex relationships among them. 
	%All in all, none of the aforementioned models can represent the various kinds of information from source code in one model. Through the practice in software engineering, combined source code models \cite{tufano2018deep} show stronger competitiveness in performance compared with other source code models. So source code representation is the key problem in the research on comment generating automatically, and seeking better source code representation models is still an open research topic.}
\begin{figure*}[t!]
	\centering
	\includegraphics[width=0.8\textwidth]{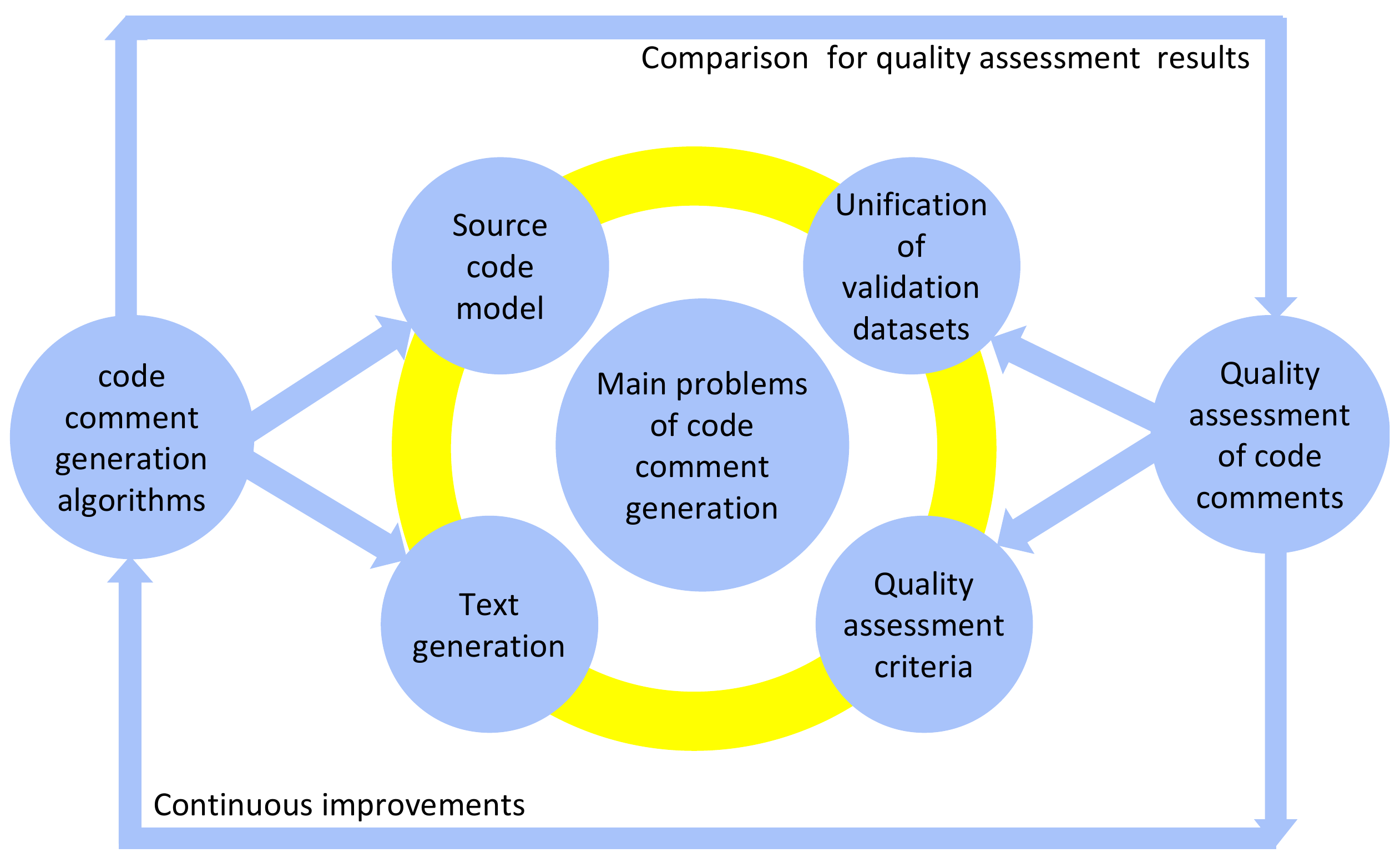}
	\caption{The framework of code commenting research.}
	\label{fig2}
\end{figure*}
	
 \textbf{Text generation.} 
	Since a natural language is unstructured and its expression form is flexible, the task of generating natural language text is difficult \cite{angeli2010simple}. When it comes to code commenting, information should first be extracted from source code accurately before constructing natural language comments, which makes it more challenging. % Moreover, extracting features accurately from source code itself poses further challenges.exists some deviation. So generating comments based on these features is certainly to produce bigger deviation. 
	%Yet, generating text is the essential step for automatic program annotation. Existing solutions could not obtain satisfying results.
	Existing solutions for text generation can be classified into three categories. First, rule-based text generation solutions, which generate text depending on designed rules or natural language models (templates) \cite{moreno2013automatic,sridhara2010towards,wang2017automatically,mcburney2014automatic,sridhara2011generating,rastkar2011generating,sridhara2011automatically,oda2015learning}. Second, generative-based methods, which yield text by decoder \cite{hu2018deep,zheng2017code,allamanis2016convolutional,mou2016convolutional,hu2018summarizing}. Finally, search-based text generation solutions, which produce natural language comments through retrieving existing comment text from corpus \cite{wong2013autocomment,wong2015clocom,panichella2012mining,vassallo2014codes,haiduc2010supporting,haiduc2010use,rahman2015recommending}. %are summarized as follows, one method is that generating natural language comments takes advantage of the language model elaborately designed; and the other method is identifying salient words to form summary comments\cite{haiduc2010supporting,eddy2013evaluating,rodeghero2014improving}; For deep neural networks based commenting system, outputting words one by one form the comments according to the input of source code. All the comments produced by these algorithms belong to explanation comments or summary comments.}

 \textbf{Challenge 2: comment quality assessment.}

Quality assessment of code comments is another key problem for code comment generation. There exist two main issues in comment quality assessment: unification of datasets used for validating and testing commenting algorithms, and selection of evaluation criteria.

%\Figure[t!](topskip=0pt, botskip=10pt, midskip=0pt){fig2.png}
%{The framework of code commenting research.}

 \textbf{Unification of datasets for verifying commenting algorithms.} 
At present, there exist many algorithms for automatic or semi-automatic code comment generation. %and we divide them into three categories and eight sub-categories (details see Section \ref{sec:algorithm}). 
These studies exploit different datasets to test their algorithms, which makes it difficult to compare testing results and performance of algorithms. As a result, unifying the datasets for testing is very important. However, because each specific comment generation algorithm has the language dependency, unification of dataset for testing is challenging.
	
 \textbf{Selection of evaluation criteria for quality assessment of code comments.} 
	
The lack of appropriate quality assessment metrics will lead to the absence of a quantitative comparison that highlights the strengths and weaknesses of each commenting algorithm. In existing work, the criteria of quality assessment of code comments are different depending on the category of comments. For example, according to \cite{der2012quality,steidl2013quality}, from the perspective of functions, comments can be categorized into descriptive comments, summary comments, conditional comments, comments for debugging and metadata comments, etc. %The criteria of quality assessment of comments are different depending on the different purposes. For example, comments for debug should be removed. % Some kinds of comments well serve for maintenance staff, for example, by mining discussion content from posts at the Q\&A websites we can acquire comments for code discussed in posts, and posts of  Q\&A websites contain varying information e.g. defects of source code, code concerns and restrictions etc., the evaluation criteria for this kind of comments are different from summary comments’s; The evaluation criteria of summary comments used in existing literature \cite{sridhara2010towards,mcburney2014automatic,sridhara2011generating} are accuracy and conciseness in content; the quality assessment criteria of code comments fitting for maintenance workers are popularity, relevance, etc., the meaning of these evaluation criteria of comments will be discussed in Section 4.
 Even in the same category, different automatic comments generation techniques adopt different comment assessment criteria. Thus it is important to design and formulate appropriate quality assessment metrics for comments, which will promote the study of automatic code comment generation.

%{This paper will classify and summarize the problems and key technologies in the studies of automatic code commenting. Through analyzing the problems mentioned above, this papre will focus on the key problems such as, the algorithms of automatic commenting and quality assessment of code comments, etc., and try to provide an overview of the recent advances on the automatic code commenting studies, and lay out a vision for future researches in this area.}

\begin{figure*}[t!]
	\centering
	\includegraphics[width=0.7\textwidth]{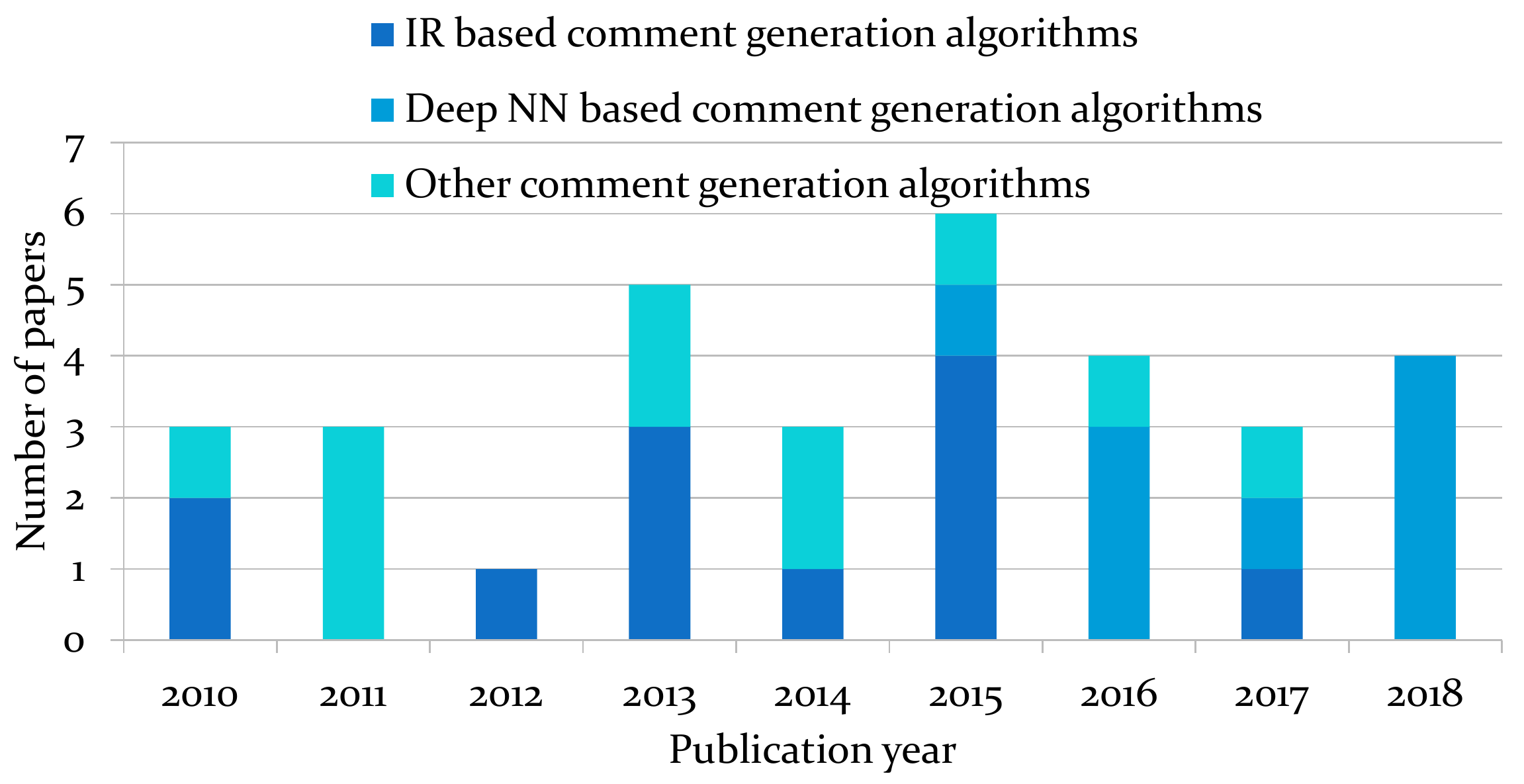}
	\caption{Distribution of the selected 32 research papers over publication years.}%Number of publications from 2009 to 2018 that witness novel research contributions and different algorithms.}
	\label{fig3}
\end{figure*}
%\Figure[t!](topskip=0pt, botskip=10pt, midskip=0pt){figc2.png}
%{Number of publications from 2009 to 2018 that witness novel research contributions and different algorithms.\label{figc2}}

\subsubsection{Research Framework}
Over years, software quality has always been an important research topic in software engineering. Quality of code comments is one of the important factors for evaluating software quality. As early as the 1980s, researchers began to study code comments. At present, the literature on the study of code comments focuses on relations between comments and the readability of code, relationships between comments and code understandability, algorithms of automatic program annotation and quality assessment of code comments, etc. In general, we summarize the studies of automatic code commenting and the related work from two perspectives:
\begin{itemize}
	\item Technologies of automatic program annotation.
	\item Quality assessment of code comments. 
\end{itemize}

{The study of automatic code comment generation techniques is a hot research topic in code commenting, and another problem associated with code commenting is the study of comments quality evaluation. 
According to \cite{b5}, there exist two kinds of comments: native comments written by code authors, and analytical comments produced
by a computer program instead of code authors. We will discuss the quality assessment of analytical comments in this paper. The other studies closely related to comment generation are those on supporting comment decision that aims at guiding developers to choose the locations needed to comment in source code. With appropriate locations, comments could well improve the readability of code. Additionally, several studies are related to source code analysis and processing, such as code suggestion \cite{hellendoorn2017deep}, generating natural language summaries for code defect \cite{rastkar2010summarizing}, crosscutting source code concerns \cite{rastkar2011generating}, class diagram \cite{burden2011natural} and source code commit \cite{liu2018neural,jiang2017automatically} etc.} %But restricted in space of the paper, the study of comment decision will not be summarized in this paper.}

In summary, this paper focuses on studies on the algorithms of comment generation and the quality assessment of comment. These two lines of work are interdependent on each other, and their relationship can be shown in Figure 2.% Algorithms of code commenting are the hottest research issues in the researches on automatic code comment generation, and quality assessment of comments is another research focus.}

\subsubsection{Trends of the development of code commenting techniques}
%{In this paper we present an analysis of the research on autocommenting conducted in the last decade.}

The research on code commenting techniques has received much attention in the last decade, fostered by the rapid spread of information retrieval, machine learning, neural networks and other related techniques. Our survey indicates that most of code commenting systems developed from 2010 to 2014 exploit information retrieval techniques, and most of code commenting systems developed in the last five years mainly adopt deep neural network techniques.%are built upon the recent advacement in deep neural network techniques.

To provide a comprehensive survey of the emerging trends in code comment generation automatically, we systematically reviewed the literature from 2010 to 2018 and select 32 representative papers from 59 papers that were published in the last ten years. %and identified as unique representatives of clusters of related publications. 
These papers focus on the main code commenting algorithms, and reflect the changing of research interests in the area of code commenting algorithms.

Note that in the process of collecting papers, we first performed two types of searches for related papers: (1) Online library search for papers containing keywords including ``code + comment'', ``comment'', ``code + summary'' and ``summary'' in the fields of title, abstract and index terms of the papers from ACM Digital Library, IEEE Xplore Digital Library, DBLP, Google Scholar and arXiv.org. (2) Specific search of major conference proceedings and journals in software engineering and artificial intelligence, including IEEE ICSE, IEEE FSE, IEEE/ACM ASE, IEEE TSE, ACM TOSEM, EMSE, AAAI and IJCAI.
Then we refined the list of the returned papers manually and read them one by one. To further collect more relevant papers on code comment generation and avoid missing important research efforts, we further performed a citation analysis on the selected papers from keywords search. A citation analysis is a manual process of reading title and abstract of candidate papers. To sum up, we selected most of the papers through keywords search, and complemented the results further by means of manual citation search. 

Figure 3 shows the distribution of papers over the years according to the types of algorithms used in the papers. The figure indicates a relevant increase of interest and results: almost half of the papers have been published in 2015-2018, and more than 60\% of papers appeared after 2014. The earliest technique used in studies on automatic code comment generation is information retrieval. In 2015, with the emergence and development of neural network techniques, deep neural network models was first applied to automatic generation of code comments. Afterwards, the interests of research in deep neural network based comment generation have been increasing dramatically over the years. At the same time, Figure 3 indicates that recent researches mainly focus on deep neural network based commenting techniques.

%% file: algorithm.tex
\section{The algorithms of AUTOMATIC GENERATION of code comments}
\label{sec:algorithm}
This section first presents the classification of code commenting algorithms, then gives a thorough analysis of principles of each type of algorithms. Finally, we summarize the characteristics of the existing algorithms. % between deep neural network based commenting techniques, a state-of-the-art comment generation technique, and other commenting techniques in terms of features and strengths and weaknesses.}

%{In the researches on automatic code commenting, the key challenge is designing the novel automatic commenting algorithms with high performance. Algorithms of code commenting are the core and key point in automatic commenting researches.}

%In existing literature, researchers propose many sorts of automatically commenting algorithms. Yet there is still no commercial automatic commenting system. The majority of studies aim at helping developers comprehend program, that is helping test staff and maintenancers understand the intents of code designer, and trying to promote their work efficiency. These commenting tools, referred to as code commenting systems, can generate natural language descriptions for source code. In general, existing researches on commenting techniques are characterized as three points. Firstly, a specific commenting tool can comment code written in a special programming language. Secondly, the code granularity that can be commented by commenting tools is different, and forms and purposes of natural language comments are different. Lastly, the quantity of generated comments is still small.

\subsection{CLASSIFICATION OF AUTOMATIC CODE COMMENTING}
%According to 30 existing high quality papers from top journals and international conferences on software engineering and data mining,
As shown in Figure 4, existing algorithms mainly fall into three categories: information retrieval based algorithms, deep neural networks based algorithms and other automatic code comment generation algorithms.% We give a tree-shaped diagram for the categories of automatic commenting algorithms and comment quality assessments researches.(see Figure 4)
 
%\Figure[t!](topskip=0pt, botskip=0pt, midskip=0pt){fig3.png}
%{The classification of the related technology of code comment automatic generation.\label{fig3}}
\begin{figure*}[t!]
	\centering
	\includegraphics[width=0.9\textwidth]{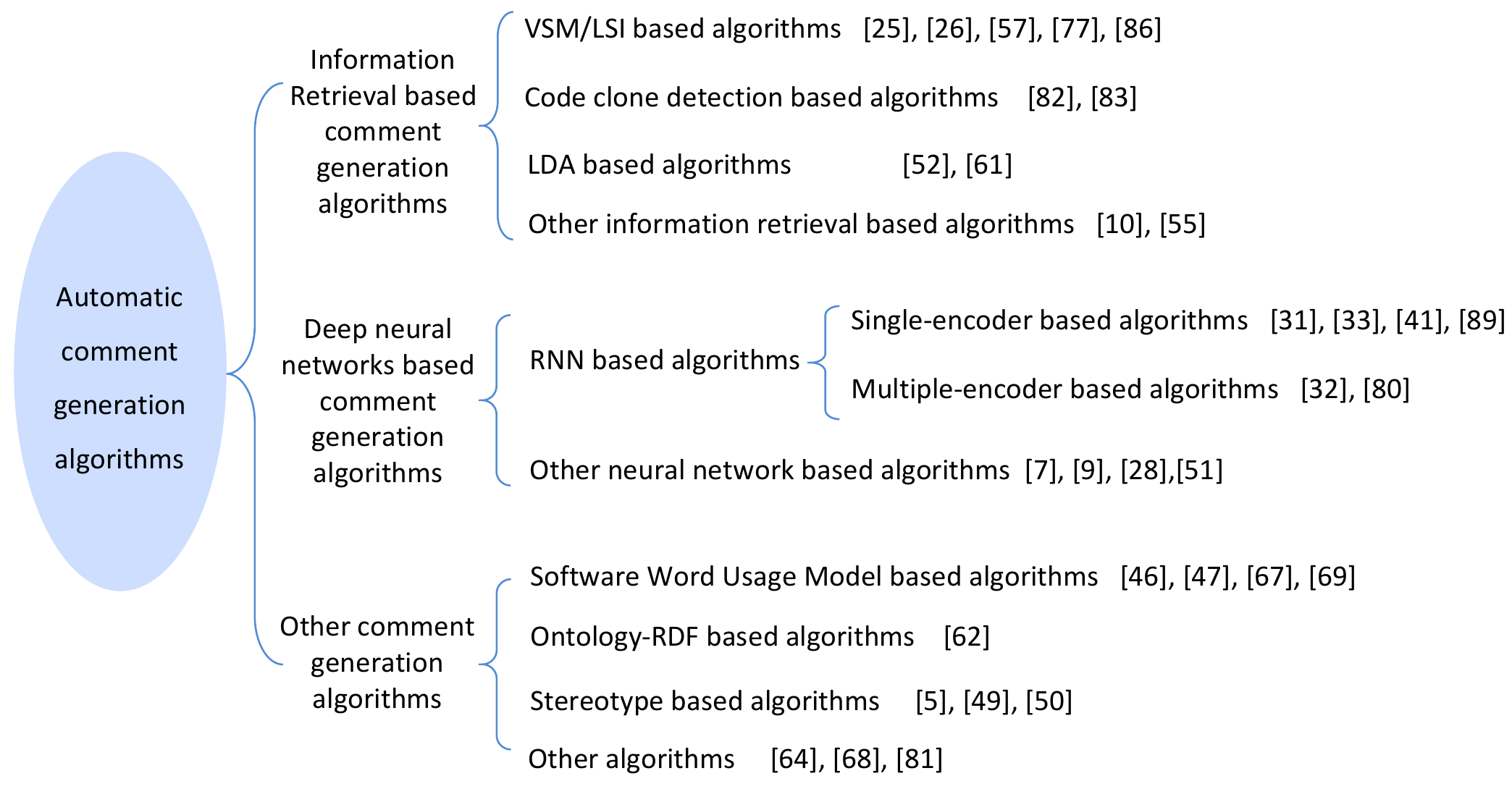}
	\caption{The classification of code comment generation algorithms and related literature.}
	\label{fig4}
\end{figure*}

In the following, we will summarize and analyze these three categories one by one in detail.

\subsection{INFORMATION RETRIEVAL BASED COMMENT GENERATION ALGORITHMS}
%Information retrieval based automatic commenting approaches refer to making use of the features of target objects-source code need to be commented, and seeking the matched objects-- source code from the prepared data set, which has the similarity in special features. Many automatic commenting systems adopt information retrieval techniques to generate comments for programs\cite{}.
 In general, given a piece of source code that lacks comments and a dataset of source code with comments, information retrieval algorithms first %construct the token-based model of target code and code segments in dataset, that is represent source code as a vector or matrix. Then 
compute the relevance between the target code and other source code in the dataset. Then one or multiple pieces of source code that well match the target code will be returned, and their comments will be used to generate the comments for the target program. The commenting algorithms based on the information retrieval techniques generally exploit the techniques based on Vector Space Model (VSM), Latent Semantic Indexing (LSI), and Latent Dirichlet Allocation (LDA) or other related techniques like code clone detection. %can be classified into two sub-categories: algebraic models based comment generation algorithms and probability models based comment generation algorithms. Essentially, the comments generated with information retrieval algorithms are based on reusing the existing comments of similar source code because no new words or sentences are produced in this process.In terms of the style through which comment produced, information retrieval based commenting techniques belong to the sort of comment reused\cite{}. Because in the process of comments generation, the generated comments do not include any new words or text, only extract the important words and text from source code in data set. 

One of the early applications of information retrieval (IR) techniques in the area of software engineering is about the traceability between code and comments. In 2003, Marcus et al. \cite{marcus2003recovering} employed the latent semantic index (LSI) technique, to analyze source code and the external documents for extracting semantic information from program and documents, and they further recovered links between documents and source code. Although the study itself is not on the problem of automatic comment generation, the proposed method can be applied to code commenting. Kuhn et al. \cite{kuhn2007semantic} use Latent Semantic Indexing (LSI) technique to find the linguistic topics which refer to the informal semantics information contained in the identifiers names and comments in source code. And these linguistic topics reflect the intention of the code, and cluster source code according to topics. 
% to solve the problem of automatic comment generation.  

In existing literature similarity comparison is not performed directly in the form of source code text. Source code usually is converted into varying representation forms that is fit for follow-up comparison. Most commenting systems convert source code into the form of parse tree \cite{wong2013autocomment} or abstract syntax tree (AST) \cite{wong2015clocom}, %\sxt{which all belong to token-based source code models}; 
then, compare the target code with other source code from datasets. %Some studies invoke designed function to compare cosine similarity of two code snippets. 
According to the comparison results, the  matched code is returned. The corresponding comments of the matched code are filtered with some heuristic rules. Finally the most relevant text description is recommended as the comments or summaries for target source code. All in all, these kinds of algorithms generally generate comment texts by searching or designed rules.

%\subsubsection{Similarity based comment generation algorithms}Information retrieval based comment generation algorithms also are referred to similarity based comment generation algorithms. %aim at checking up the matched source code for target code from dataset, which consists of <comment, source code> pair, and extract the corresponding comments of matched code fragments or other forms of natural language descriptions, then output them as comments for target code. 

Some commenting systems \cite{wong2013autocomment,wong2015clocom} require high quality datasets which contain source code and the corresponding comments pairs in order to generate comments for programs. %We need many high quality datasets (e.g. source code files which consist of source code and its corresponding comments, or discussion information between developers, %which include code comment and the related discussion text,} 
%and need to be processed further for use), so that commenting systems can recommend high quality comments for target code.
Some other researches make use of code clone detection techniques to find matched source code for target code .

\subsubsection{VSM/LSI based Comment Generation Algorithms}
These kinds of algorithms refer to leverage Vector Space Model (VSM) \cite{b6}, Latent Semantic Indexing (LSI) \cite{b2}, or combination of both to generate comments for classes and methods.  %Some commenting algorithms adopt code clone techniques to find out the similar code segments. We divide this algorithms into algebraic model based algorithm. %and further provide textual description for target code. The reason is that they all exploit vector, one of the algebraic representation, to represent source code. %and other source code comment generation algorithms which exploit 

VSM and LSI both belong to the techniques of information retrieval. They initially developed for the tasks of natural language processing. %VSM represent source code text and query text as is used to judge whether two documents are similar according to the quantity of words that commonly appear in both documents \cite{b6}. 
When we employ VSM and/or LSI to generate comments for source code, source code text or query text are usually represented as vectors, matrix or tuples \cite{hu2018summarizing}. Each element in the vector denotes the weight of a word in the documents. There are many methods to compute weight of term in VSM, and term frequency-inverse document frequency (tf-idf) is the most widely used weighting methods. Employing singular value decomposition (SVD), LSI recognizes term relevance between terms and concepts, and extracts the conceptual topic of a text. The commenting system determines whether the term should appear in the comments of source code according to the value of weight for each term, or calculates text similarities between the vector of query text and source code text. %and represent similarity as scalars. By sorting similarities of words,
The term with higher similarity represents higher relevance to the code snippets or the topic of queries. So commenting systems recommend these terms as the key words to construct comments for target source code. 

These techniques are used to mine code text, and find out key words in source code text for constructing natural language description as code comments. These comments are used to describe the functions, characters and parameters of some form of source code, such as classes, methods or code block etc.

Summaries of source code can be treated as the special comments which can help developers and maintenance engineers comprehend programs more accurately and faster, saving their precious time. Several years ago, researchers have tried to summarize source code by exploiting VSM and LSI techniques. The achievements have been presented in the literature \cite{haiduc2010supporting,haiduc2010use,vassallo2014codes,ying2013code}. 
 
% \textbf{a) VSM/LSI based comment generation algorithms.}
 Haiduc and Aponte et al. \cite{haiduc2010use} use VSM and LSI to analyze the source code text, and generate the extractive and abstractive natural language summaries for classes and methods. First, they convert source code documents and packages into a document collection, called corpus. Then they represent the terms which are included in the identifier names and comments from source code and documents in the corpus as a matrix. %where each row represents a term and each column represents a document.The content of a cell in the matrix represents the weight of a term with respect to a document. 
 When generating extractive summaries for source code with VSM, the most relevant terms in the source code document are selected according to the chosen weight. %, and the top K terms are included in the summary comments. 
 At the same time, they also use LSI techniques to calculate the cosine similarities between the vector of each term in the corpus and the vector of a source code document to be summarized, then generate terms with high similarity %are sorted according to their similarity values, and the top n terms are included in the summary comment. The feature of this approach is that it can contain terms 
 that do not appear in the method or class to be summarized, but appear in the corpus. In this way, they analyze the method and class source code in Java project and generate short and accurate textual descriptions for them. In another effort, Haiduc et al. \cite{haiduc2010supporting} exploit LSI only to generate summary comments for the code of Java class in open source repository. 
 
Exploiting the same approach, Vassallo et al. \cite{vassallo2014codes} use VSM model to represent source code text and developer discussion text from Question and Answer (Q\&A) on Stack Overflow as vectors, and calculate the cosine similarity between target source code text and discussion text to find the maps. The paragraph texts with high similarity are recommended as the comments of target source code. As a result, they mine the crowdsourcing knowledge to recommend comments for method.
 
Similarly, Panichella et al. \cite{panichella2012mining} employ heuristic and Vector Space Model to process and analyze developer communications for methods descriptions. The developer communications mainly refer to emails and bug reports that are related to classes, methods and parameters. They extract paragraph texts which can be traced to source code methods and recognize the relevant paragraph by means of computing textual similarities (that is cosine similarity) between text paragraph and the text of each traced method. The relevant paragraph with high similarity is recommend as the method description.  % signatures' description as the comment for target programs. They find out the related text paragraphs by filtering emails and bug reports, and index the paragraphs and methods using a Vector Space Model. %And they calculate text similarities between text paragraphs and source code text of methods. According to sorting results on similarities, recommend text paragraph with high similarity value as the comments of target project. 

The drawback of this type of technologies is that it only takes into consideration of terms that appear in the corpus or source code documents, and pays no regards to other information that is contained in source code documents, e.g., program invocation, data dependency, words sequence in source code. Therefore, it is difficult for these systems to improve the accuracy of generated comments further.
%{Haiduc et al. exploit automated text summarization techniques to yield short and accurate textual description for source code. They extract identifiers and comments from source code, and create source code corpus after processing further. Then they use text retrieval techniques, VSM and LSI, to generate summaries for source code. Concretely, firstly represent the terms and documents in the corpus in a matrix, then, }\begin{itemize}
%    \item \textbf{Crowdsourcing knowledge based commenting systems}

\subsubsection{Code Clone Detection based Comment Generation Algorithms}
Code clone detection based comment generation algorithms are concerned with utilizing code clone detection technique to find similar code in a database, and the corresponding comments of the matched code or discussion text are viewed as comments for the target code.

Wong et al. \cite{wong2013autocomment} propose an approach based on code clone detection techniques, which mines comments from a large programming Question and Answer (Q\&A) site. They mine posts from Q\&A of Stack Overflow, where developers can post questions and receive the corresponding solutions in Q\&A site. The posts from Q\&A on Stack Overflow, containing code snippets together with the corresponding textual descriptions, are referred to as code-description mappings. They extract such mappings to build a database. Then they leverage code clone detection technique, i.e. the longest common substring, to discover similar code segments in database, and extract corresponding comments for target code segments. In a different effort, Wong et al. \cite{wong2015clocom} mine comments from open source software projects from software repositories in GitHub. With the help of improved code clone detection techniques they find out more matched code snippets. Their new code clone detection tool leverages code parser to build ASTs for code snippets, and tokenize the serialization of AST node of code before comparison. In this way, their clone detection tool takes into consideration of structural information of source code so as to find out more matched code snippets. As a result, the improved code commenting system generates more comments for target code. % than the previous one. However, the quantities of generated comments are still small in the whole.made improvements in two main points. First, they selected open source software projects from software repositories in GitHub instead of the posts of Q\&A site in Stack Overflow as the high quality datasets which containing code and the corresponding comments. Second, they adopted improved code clone detection technology. }%Here, the code clone detection technique calculates the longest common substring of code segments between the databases and the target software to detect similar code segment. However, 

The principles of code clone detection based comment generation algorithms are simple, and the quantity and quality of the generated comments heavily depend on the scale and the quality of the dataset built for commenting systems. Consequently, when we require to provide comments for source code written in particular programming languages in certain area, it is important to build a high quality dataset containing code and comment pairs written in the same programming languages in the same domain.

The quality of comments generated by these commenting systems, to a large extent, depends upon the performance of code clone detection algorithms and the quality of comments from datasets. Accordingly, to improve the quality of the generated comments, it demands we collect more high quality open source software projects or more discussion and communication information which contain code snippets and the corresponding natural language comments. %On the other hand, it is beneficial to the comment generation algorithms that improve the code clone detection algorithms in the accuracy.

The drawback of this approach is that the quantity of the generated comments is much smaller. The reason is that the quantity of the generated comments heavily depends on the information contained in databases or the open source software projects from GitHub. For example, if a code segment is never discussed in any posts, commenting system will fail to recommend any comment for it at all. % which is one of websites widely used to ask questions about code development and debugging, etc.
% \item \textbf{Commenting systems leveraging high quality open source software projects}
% \item \textbf{Mining code description from the communication discussion of developers}
%\end{itemize}

\subsubsection{ LDA based Comment Generation Algorithms}
Latent Dirichlet Allocation (LDA) \cite{blei2003latent} is one of the topic models proposed for automatically extracting topic from text documents. It is one of the probability models. Probability models used in comment generation algorithms include n-gram language models, Latent Dirichlet Allocation (LDA) \cite{blei2003latent} and the other variants of LDA. N-gram models are widely used in statistical natural language processing. LDA is an IR model that can fit a generative probabilistic model from the term occurrences in a corpus of documents \cite{panichella2013effectively}. LDA based comment generation algorithms are concerned with building the source code model with LDA model and generate comments for target source code. %by fitting the generative probability distribution from the token of source code occurrences in a corpus of document of source code. 
In other words, LDA can extract particular features of source code. When N-gram model is used to solve automatic comment generation problems, it is usually used to assist other statistical model to analyze source code, or train source code models. Its model is simple and effective, and it becomes the effective model for semantic mining from source code documents.
%\textbf{a) LDA based commenting algorithms}

In another effort, Movshovitz-Attias and Cohen  \cite{movshovitz2013natural} use topic models, LDA, and n-gram models to predict comments for Java source code. They train n-gram models and LDA models over the same source code documents from multiple training datasets respectively. Then they consider documents as having a mixed member of two entity types, code and text tokens, and train link-LDA models over the documents. Using trained models they compute the posterior probability of document topics and with which they further infer the probability of the comment tokens. Finally the comments tokens with high probability are recommended as comments for source code files \cite{rahman2015recommending,movshovitz2013natural}.

Employing LDA, Rahman et al. \cite{rahman2015recommending} analyze discussions from Stack Overflow Q\&A site to recommend insightful comments for open source project. They exploit a heuristic-based technique, which is different from Wong\'s \cite{wong2013autocomment}, to mine the crowdsourcing knowledge to yield comment for open source project. The generated comments mainly describe the deficiency, quality and scope of source code to improve source code and can help maintenance engineers to perform the maintenance tasks.%build topics model for each API corpus using LDA. After acquiring the probability distributions in the corpus, and check text similarities between statements from open source projects and the code extracted from posts in Q\&A community. When the matched code was found, the topics of the corresponding user comments and answers are selected to act as } 

\subsubsection{Other Information Retrieval based Comment Generation Algorithms}
Oda et al. \cite{oda2015learning} use phrase-based machine translation (PBML) and tree to string machine translation (T2SMT) to generate pseudo-code from Python source code. PBML and T2SMT are both statistical machine translation frameworks. Pseudo-code is a natural language description, and an informal high-level description of the operating principle of a computer program \cite{b7}. Pseudo-code can aid novice developers or inexperienced readers to understand programs, so it can be treated as a special type of comments. First, they construct the source code/pseudo-code parallel corpora (a dataset, which consists of <code, pseudo-code> pairs), that is, they add pseudo-code for existing source code by human labors. Then they exploit two frameworks (PBMT and T2SMT) and some existing, open source tools to train the pseudo-code generator over the parallel corpus. Here, open source tools include: tokenizer of the target natural languge, tokenizer and parser of the source programming language. Through training they acquire the translation rules automatically, and finally, the trained pseudo-code generator can output NL pseudo code for target source code.

In a different effort, Allamanis et al. \cite{allamanis2015bimodal} exploit probabilistic models that jointly model short natural language expressions and source code snippets, which represent the mapping relation between source code snippets and the corresponding natural language queries. They train parameters of their model which denote mappings between source code snippets and natural language queries. After building the model, which allows mapping in both directions: from natural language to source code, and from source code to natural language, this model can be used into two retrieval tasks, i.e. retrieving source code snippets for a natural language query, and retrieving natural language descriptions for a source code query. For the second task, matched natural language descriptions can be regarded as comments of a given source code. %Allamanis et al. \cite{allamanis2015bimodal} exploit vectors to represent tokens of parser trees of source code and natural language queries. The multiplication of two vectors represents the corresponding joint model.}
 
To sum up, information retrieval technology is the earliest trial for researches on code comment generation. With the development of artificial intelligence and machine learning technologies, researchers continually apply the emerging techniques to automatic comment generation researches, such as deep learning algorithm. We will summarize and compare this type of technology with others in detail in the following subsection.

\subsection{DEEP NEURAL NETWORKS BASED COMMENT GENERATION ALGORITHMS}
%After 1986 the researches on machine learning and deep neural network are coming out of trough and gradually recovering, and researchers have studied several years and found that
Recent years have witnessed deep neural networks' excellent performance on natural language processing, machine translation, image recognition and speech processing \cite{kalchbrenner2014convolutional,sutskever2014sequence,rush2015neural,vaswani2017attention,cho2014learning}. In the field of software engineering, researchers formulate the conversion task from source code to comments as a translation problem between programming languages and natural languages, and they try to exploit deep neural network approaches to solve source code commenting problems \cite{hu2018deep,zheng2017code,allamanis2016convolutional,allamanis2015suggesting,mou2016convolutional,iyer2016summarizing,wan2018improving,hu2018summarizing,liang2018automatic}.%Some studies treat source code as the plain text, and leverage neural networks to summarize source code, and generate summary comments for source code\cite{rush2015neural}.

Deep neural network based comment generation algorithms fall into two main categories: RNN based algorithms and other neural network based algorithms. There are three kinds of deep neural networks: Convolutional Neural Network (CNN), Recurrent Neural Network (RNN) and Recursive Neural Network (RvNN) \cite{pouyanfar2018survey}. Convolutional neural networks are fit for solving the problems of natural language processing (NLP), image recognition and speech processing \cite{angeli2010simple,kalchbrenner2014convolutional}. Recurrent neural networks are good at processing and predicting sequential data, and are preferred in NLP and speech processing. So, RNN is fit for generating comments for source code. RvNN is preferred in NLP too, and it can process source code to generate comment for code \cite{liang2018automatic}.% We classify the CNN based commenting algorithms into other neural network based comment generation algorithms. %\sxt{In the respect of source code models, most of these system adopt token-based model or syntax-based model, that is they represent source code at the level of abstract syntax tree of source code. And these kinds of algorithm generate comments by decoders, and we call it generation-based approach. The details of principals of deep neural networks are listed as follows.
When borrowing neural machine translation techniques as the solution of commenting problems, two important structures, encoder-decoder structure and attention mechanism, are worth mentioning. They are usually used in deep neural networks based commenting systems to assist to generate comments for source code.  

\textbf{Encoder-Decoder framework}
In deep neural network based comment generation systems encoder-decoder structure, also known as sequence to sequence model, is generally exploited. In the structure of encoder-decoder, the encoder plays the role of encoding source code into a fixed-sized vector; and the decoder is responsible for decoding source code vector and predicting comments for source code. % The structure of encoder-decoder model is shown in Figure 7. 
The difference among various encoder-decoder structures lies in the form of inputs and the type of neural network \cite{gehring2017convolutional}. Generally, the inner structure of encoder-decoder can choose RNN, CNN \cite{gehring2017convolutional} and the variants of RNN, such as Gated Recurrent Unit (GRU) and Long Short term Memory model (LSTM).

%\Figure[t!](topskip=0pt, botskip=20pt, midskip=0pt){fig4.png}{The structure of the Encoder-Decoder model\label{fig4}}

%{From the structure of Encoder-Decoder in Figure 5, we can see that the unique relationship between encoder and decoder is a fixed-sized semantic vector $c$, that is to say an encoder compresses the whole information from input sequences into the fixed-sized semantic vector $c$. The problem of these structures is that a single semantic vector cannot represent full information of the input sequence roundly, because that the last input content will override the former input content. In this way, with the length of input sequence increase, especially the length of input sequences longer than the length of source code sequences from training dataset, the performance of decoder is becoming poor. In order to solve this problem, Bahdanau et al.\cite{mnih2014recurrent} propose a new attention mechanism.}
\textbf{Attention mechanism}
Attention mechanism is usually added to the encoder-decoder framework. It is responsible for dynamically assigning the higher weight values to more relevant tokens of each word in the input sequences of decoders. In this way, it makes the effectiveness of decoders improved. Attention mechanism is proposed by Bahdanau et al. \cite{mnih2014recurrent,bahdanau2014neural}. It is a good solution for the problem of the poor performance in the case of the long sequences. %Attention mechanism is added to the Encoder-Decoder model, this mechanism computes similarity between input sequences and output sequences first, then assigns according to the similarity,  The more similar between the state of current input and target state, the weight value of the current input words is bigger, then the current output depends on the current input larger. So the role of attention mechanism is that when decoders process the input sequence to produce the words of output sequences, the decoders would take more considerations on the important words in input sequences. The attention mechanism is adopted extensively in commenting systems based on the machine translation techniques and machine translation systems\cite{luong2015effective}.  

Since Deep neural network based comment generation algorithms belong to the categories of machine learning, commenting generation systems based on the Deep Neural Networks require high quality datasets containing code and comments to train the neural network. The dataset can meet all data requirements for the system, providing data for training and for validation and test of the commenting algorithm as well. 

When training and adjusting the parameters for the neural network model, we
could take some preprocessed source code, (such as the AST of the code or the sequences of AST of source code, etc.) as the input data. The corresponding comments of source code can be treated as the output of the commenting system. Trained neural networks can generate natural language descriptions for target programs. 

\subsubsection{RNN based Comment Generation Algorithms}
In order to model temporal sequential information with neural network, developers design RNN algorithm. RNN is a popular and widely used algorithm in deep learning \cite{cho2014learning}, especially in natural language processing and speech processing \cite{li2015constructing}. %The basic structure of RNN includes three layers: an input layer, a recurrent hidden layer and an output layer. Where, an input layer represents each word to a vector, and a recurrent hidden layer recurrently computes and updates a hidden state after reading each word, and an output layer estimates the probabilities of the following word given the current hidden state. 
When RNN algorithm servers as the solutions for comment generation, the encoder-decoder structure and attention mechanism are commonly used so as to improve the accuracy of commenting systems. %Generally, the RNN encoder-decoder model uses two RNNs: one encode source code tokens sequence into fixed-length vectors, and the other RNN decodes the vector into natural language comments. %\sxt{In this process, attention mechanism introduces a dynamic length vector between encoder and decoder, and jointly learn the alignment of input and output sequence.
 %The network structure of neurons in RNN is displayed in the Figure 5. The left of Figure 5 depicts the structure of RNN in the whole, the right of Figure 5 represents the classic structure of RNN. RNN is distinctly different from CNN in the network structure, which increasing interconnections and information transferring between the neurons in hidden layers, and this makes the output of neuron to act on itself at the next timestamp, that is, the input of neuron in the i layer at the time of t include the output of itself in the time of m-1, in addition to the output of neuron of i-1 layer.

%\Figure[t!](topskip=0pt, botskip=0pt, midskip=0pt){fig5.png}
%{The architecture of Recurrent Neural Network.\label{fig5}}

%In order to facilitate analyses, people usually unfold the structure in Figure 5 to the structure of Figure 6 in temporal sequence. In Figure 6, the output O$_t$ of neuron at the t time is not  only influenced by the input x$_t$ in the t time, but also influenced by the history of hidden states, such as the time of (t-1),(t-2) etc. As a result, the architecture of the RNN could implement the task of temporal sequences.
Another two important variants of RNN are Long Short term Memory model (LSTM) \cite{rush2015neural,tai2015improved} and
Gated Recurrent Unit (GRU) \cite{cho2014properties}. LSTM is a special kind of RNN, capable of learning the long term dependency information. %The LSTM was proposed by Hochreiter and Schmidhuber at 1997, and was improved and popularized by Alex Graves. LSTM succeeded at solving the problems of long term memory control in the RNN. 
The characteristics of LSTM are that it has a three-gated-controller structure and construct a controlled memory neuron which solves the gradient descent and gradient explosion in the traditional RNN. Compared with LSTM,
GRU is simple in structure, and overcomes the shortcomings of LSTM: complex structure, complicated realization and lower execution efficiency. GRU only uses two gates: one is the update gate, and the other is the reset gate. The characteristics of GRU are that it is easy to realize, having shorter runtime, simpler parameter training and more easily train than LSTM, especially in the case of processing large training data. So GRU is adopted in many applications.

According to the number of RNN used in encoder, we divide the RNN based comment generation algorithms into two categories: single-encoder based commenting algorithms and multiple-encoder based commenting algorithms.    

\textbf{a) Single-encoder based comment generation algorithms}

In single-encoder based comment generation algorithms, the encoder is composed of one RNN structure. This is a classical encoder-decoder structure used in comment generation tasks.

Iyer et al. \cite{iyer2016summarizing} propose an LSTM based comment generation model, named CODE-NN, which uses Long Short Term Memory (LSTM) networks with attention mechanism to produce natural language summaries for C\# code segments and SQL queries. The character of CODE-NN is adding the attention mechanism to LSTM model. CODE-NN is trained in the dataset automatically collected from Stack Overflow which includes title and code segment pairs. %It uses the RNN with end to end  to implement the tasks which usually require two steps, content selection and text generation, to finish. 
During the process, attention mechanism highlights the relevant and important tokens in source code, completes the relevant and important content selection, %by dynamically assigning higher weight values to more relevant tokens of the input code segments, 
and LSTM provides the context for words. After training the parameters of embedding matrices and other matrices, Iyer et al. leverage the trained model and an input code snippet to generate the natural language summaries for the code.

GRU is also used in the encoder-decoder framework to generate comments for source code. In a different effort, Zheng et al. \cite{zheng2017code} use the encoder-decoder structure whose basic element is GRU. Their attention module in encoder is a global attention mechanism. They take the embedding symbols (e.g identifiers) in code snippets as learnable prior weights to evaluate the importance of different parts of input sequences. After sorting identifiers in code segments based on the order of appearance, they encode tokens of source code into an embedded token. %They take the  so that attention mechanism is able to utilize the specific features. Their method borrows the neural network language translation model in natural. In this system,  It sorts identifiers, encodes tokens of source code, then computes the context vector, so that it is able to exploit the structure in code. Finally with the help of attention mechanism, 
In this way, their attention mechanism is able to focus on these specific features in programs. In other words, their attention mechanism can understand the structure of code better. Finally attention mechanism helps to improve the accuracy of generated comments in commenting system. 

In another effort, Hu and Li \cite{hu2018deep} suggest that the AST sequences of source code generated by Structure-Based Traversal (SBT) often bring about much structural information from source code. So they adopt sequence to sequence model and attention mechanism to generate comments for source code. In their commenting system, named DeepCom, they use  the AST sequences of source code generated by Structure-Based Traversal (SBT) as the input of neural network, and LSTM as the basic element of encoder and decoder. After training, the neural network can learn the structural and semantic information from source code, which finally generate more accurate comments for the code.% than previous models.

Since Recursive Neural Network (shorting for RvNN) can be used to represent parse trees of natural language sentences, Liang et al. \cite{liang2018automatic} apply a RvNN over the parse trees of source code to combine the semantic and structural information from the code into representation vectors. Then they leverage a recurrent neural network decoder (Code-GRU) to decode these vectors. The overall framework generates text descriptions for the code with accuracy higher than other learning based approaches such as sequence-to-sequence model. Their algorithm mainly generates summary for code blocks.
%could capture some context features in the source code, and it has the capability of memory, so RNN models are good for source code comment generation. 

\textbf{b) Multiple-encoder based comment generation algorithms}

These kinds of comment generation algorithms contain more than one encoder in the encoder-decoder structure. Because each encoder represents and extracts one type of information from source code, these comment generation algorithms can produce code comments with high accuracy with the help of multiple encoders.

Hu and Li \cite{hu2018summarizing} exploit the transferred knowledge acquired from the process of automatic API summaries to solve the problem of automatic source code comment generation. Thus their commenting system is equipped with two encoders: an API encoder and a code encoder. This system produces higher quality summaries for source code than other code summary generation systems. The core of their commenting algorithm is putting an API encoder into the RNN encoder-decoder model. %then the API encoder read API sequence into a vector. During training, API summarization model obtains more transfer API knowledge. 
With the help of learned transferred API knowledge, the RNN decoder integrate attention information collected from both two encoders to produce the summary for target code. Their algorithm outperforms CODE-NN, a state-of-the-art code summarization approach at that time. The study is innovative for adding another API encoder which enhances the performance of code summary generation. 

Exploiting reinforcement learning, Wan and Zhao \cite{wan2018improving} improve automatic code summarization. They employ one LSTM model to represent the sequential information of code, and another AST-based LSTM model to represent the structure of source code. That is, they use two encoders in the classical encoder-decoder framework. Under this framework, they exploit reinforcement learning model to solve two issues: exposure bias and inconsistency between train and test measurement \cite{keneshloo2018deep}.  %To overcome exposure bias, they use deep reinforcement learning, that is, 
They leverage an actor network and a critic network to jointly determine the next best word at each time step. %The actor network provides the confidence of predicting the next word according to current state. And the critic network evaluates the reward value of all possible extensions of the current state and can provide global guidance for explorations. 
Besides, they employ an advantage reward composed of BLEU metric to train both networks. In this way, % The main features of their study are exploiting methods from reinforcement learning and leveraging AST-based LSTM as decoder, and another LSTM network to represent the structure of source code, and adopt hybrid attention layers to combine the code structure and sequential sequence information together. 
they directly use the characters from comment assessment to optimize code summarization.

To sum up, RNN can utilize the sequential information in the input data \cite{pouyanfar2018survey}. This property is essential in comment generation where the structure embedded in the source code sequence conveys useful knowledge, such as the structure of program and lexical and syntactic information of source code. RvNN also can make predictions in a hierarchical structure using compositional vectors \cite{pouyanfar2018survey}. %take a recursive data structure of source code and generate a compositional vector representation . 

\subsubsection{Other Neural Network based Comment Generation Algorithms}

Although seq2seq model generally adopts RNN based encoder-decoder structure, CNN model also can be used to construct encoder-decoder architecture \cite{gehring2017convolutional} in the task of natural language translation. The strength of CNN encoder-decoder is that the computation over all elements can be fully parallelized during training to better exploit the GPU hardware. CNN model can be used to represent the syntactic and semantic information of source code. With the assistance of the special attention mechanism or the RNN decoder, CNN can solve the problem of input sequence\cite{gehring2017convolutional}.  %e.g, language translation and code summarization. 
When it is applied to sequence problem, the convolution model can represent %compositional structure over sequences of input and represent
input sequential data hierarchically. The role of CNN lies in that it can extract hierarchical feature representation from input sequence. %and compute parallel. As mentioned above, to model source code with CNN, we usually import the attention mechanism into commenting system, and attention features can represent the position of each word in source code and represent this kind of knowledge into feature vector. In this way, under the help of \subsubsection{Combined Neural Network Based Comment Generation Algorithms}

Allamanis et al. \cite{allamanis2016convolutional} introduce a convolutional network into attention mechanism, which can produce short and descriptive summaries for source code snippets. The commenting system adopts the encoder-decoder framework and attention mechanism, where the basic element of decoder is GRU. Their convolutional attention module applies convolution on the input source code snippets to detect %local time-invariant and long-range topical attention features in a context-dependent way. The role of convolutional layers in this model is to detect
patterns and determine the important token sequences where attention should be focused. %The convolutional attentional mechanism is used to detect patterns and identify important tokens in the input even if they do not appear in the training set, and decoder will copy these tokens directly to the output sequences. Here, the GRU is used to solve the long-term preservation for important information in the system.
So generally CNN models are not commonly used to model temporal sequences, apart from those mentioned in \cite{gehring2017convolutional}. We usually select CNN model and other auxiliary model to solve the temporal sequences problems.

In another effort, Mou and Li et al. \cite{mou2016convolutional} exploit convolutional neural networks to represent the abstract syntax tree of source code as vectors, then classify programs according to functionality. They take the entire AST of a program as input, and design a convolution kernel over AST of source code to capture structural information of source code. Although the goal of their work is not to generate comments for programs, their source code model can be used for generating summary for program in future work.

%Based on neural network, Chen et al. \cite{hellendoorn2017deep} build the Bimodal Variational AutoEncoder (BVAE) model about source code and natural language query. BVAE model builds source code model and natural language model separately, the  encoder with Multilayer Perception (MLP) structure and RNN-based decoder jointly finish the code summarization task, the every neuron in decoders equips with GRU structure. In this system, researchers still exploit Beam Search structure to improve the summary quality.

Although most of the aforementioned approaches adopt encoder and decoder structure, Allamanis et al. \cite{allamanis2015suggesting} adopt a log-bilinear model in neural network to suggest method and class names. Their model leverages continuous embedding to represent identifier name. They believe identifiers with similar vector will appear in similar contexts. So they recommend the name of class or method by selecting the name with similar vector, and that means comparing the vector between the function body and candidate identifier name. In a word, they exploit a neural probability model instead of encoder-decoder framework to solve the method naming problem.      

%Researchers usually choose RNN models, not CNN models, to solve the source code commenting problems. The reason will be discussed as following. 
In existing literature, RNN and CNN all can be exploited to model source code and generate natural language summaries or comments for source code \cite{allamanis2016convolutional,mou2016convolutional}. RNN makes use of GRU or LSTM to represent the long distance features between long input sequences. CNN exploits a convolutional attention or convolution layers to collect and represent the features and position model of source code.

All in all, in the deep learning based commenting algorithms, RNN, LSTM, GRU or CNN are generally adopted to model source code and perform various SE tasks. % The reason is that CNN is not fit for solution of temporal sequences, if you exploit CNN to generate summary, often require other auxiliary process. Yet RNN(e.g. LSTM) usually is suitable to dealing with the tasks or problems relate to temporal sequences.
In comment generation models, RNN and CNN models are equipped with attention mechanism to improve the accuracy and efficiency. Although deep neural network based comment generation algorithms receive successes in comment generation, they still have difficulties in modeling multiple complex information from source code at the same time. Combined model is a popular direction in future researches. 

\subsection{OTHER COMMENT GENERATION ALGORITHMS}
%Other comment generation algorithms are concerned with the commenting algorithms that exclude information retrieval techniques and deep neural network algorithms.

We further present some other technologies for generating code comments. These technologies mostly either adopt some existing models from other research areas to represent source code or adopt models that exclude the aforementioned models, such as Software Word Usage Model (SWUM) \cite{hill2010integrating}, Ontology based Resource Description Framework (RDF) \cite{rastkar2011generating} and stereotype identification \cite{dragan2010automatic} etc. These models can represent the structure and
semantics of source code. % lexical, semantic, grammatical and information about program structure, program invoking, data dependencies and context information etc., and play an important role in improving accuracy of commenting systems. 
Many commenting systems \cite{sridhara2010towards,mcburney2014automatic,sridhara2011generating,mcburney2016automatic} apply software word usage model (SWUM) to represent structural, semantic and lexical information in source code; some systems \cite{rastkar2011generating} exploit ontology based Resource Description Framework (RDF) to depict semantic information of source code, and use heuristic method to find out key facts in the source code, etc. Several efforts \cite{moreno2013jsummarizer,moreno2013automatic,abid2015using} leverage stereotype identification techniques to generate summaries for Java classes and methods. %Several systems directly analyze existing source code, and make out the rules artificially, or apply heuristic to recognize features of program and context information etc. 
Finally these solutions improve the accuracy of comments and performance in commenting systems to some extent.
\subsubsection{Software Word Usage Model based Comment Generation Algorithms}
Software Word Usage Model (SWUM) was proposed by Emily Hill in 2009. She designs this model in \cite{hill2009automatically,hill2010integrating}. %which is inspired by the analogous model, 
This model is a new representation model for source code. Compared with software BOW (short for Bag Of Word), which is the commonly used model for software in earlier time, SWUM represents much textual and structural information in source code to improve software searching and program exploration. SWUM combines textual and structural information of source code into one model.

Hill proposes SWUM model to represent the facts in source code. An SWUM model is composed of three layers \cite{hill2010integrating}: SWUM$_{word}$ which models program words, SWUM$_{program}$ which models program structural information, and SWUM$_{core}$ which models phrase structure in source code, and bridge edges connect different layers. Thus SWUM can not only extract the lexical and structural information from source code, but also links between the linguistic information and structural information. With this model, developers have designed different score functions in order to complete the different application tasks. According to score values calculated by score functions, the focal phrases and key structural information are selected from source code. Finally with these focal phrase and key words, comments are produced with the assistance of designed language templates.

Sridhara and Hill et al. \cite{sridhara2010towards} utilize SWUM to represent source code and generate comments automatically for Java methods and the focal parameters of java classes. Through traditional program analysis and natural language analysis they first construct SWUM, then select the content to be included in the summary comment, and leverage the focal terms and keywords to construct the natural language text from templates for method. After combining and smoothing the generated text, the summary comments are yielded. Similarly, Sridhara and Pollock et al. \cite{sridhara2011generating} combine SWUM and heuristics to produce parameter comments for method and add them to the method summaries. They first leverage structural and linguistic information of source code to determine the content to be included in summary of parameter of method, then generate succinct description phrases with the assistance of templates. These description phrases are used to describe the intent of method. %Although their approaches are successful at initial steps toward automatic comment generation, they have two main limitations. First, the techniques can only generate comments for specific code structures (such as, one method body, or groups of method calls). Second, performance depends upon high quality identifier names and method signatures. Poor names of identifiers and methods may cause the commenting algorithm to fail to generate accurate comments or any comments at all.}
	
Based on the same model, McBurney et al. \cite{mcburney2016automatic} combine PageRank algorithm and SWUM to generate summary for Java method. They use SWUM to represent source code and extract keywords about the behavior of important methods. The role of PageRank is to discover the most important methods in given context. Their summary is of higher quality owing to the addition of context information of method. Here, the context information of method is obtained by the analysis of method calls. %The task of analysis of method calls is performed by PageRank algorithm.  %epresent source code. At the same time, they utilize  to discover the most important methods in the given context and extract the focal statements and keywords from SWUM of source code. Through adding contextual information into the summary templates, it makes the summaries more accurate. 
These commenting systems all adopt SWUM to represent source code models.%In the process of building SWUM, MeBurney et al. exploited structures as, the AST of source code, control graphs, etc.}

\subsubsection{Ontology-RDF based Comment Generation Algorithms}
Rastkar and Murphy et al. \cite{rastkar2011generating} propose an approach that automatically produces a natural language summary for crosscutting source code concerns \footnote{Crosscutting source code concerns: are concerned with source code that crosscut the modules defined in the code.}. They extract the structural and lexical information from source code and exploit ontology instance to store and describe the extracted semantic information, and they manipulate an ontology instance through an RDF graph which is used to depict resources originally. In their approaches, triples (resource, attribute type, attribute value) are used to represent one attribute of a particular class. Then they use a set of heuristics to find salient code elements and patterns from the related code concern. Finally, the information extracted in the previous steps is used to construct summary comments according to templates.  %A triple in resource description correspond to a natural language sentence, the resource in triples corresponds to the subject in natural language sentence, the attribute type corresponds to the predicate of sentences, the attribute value express the object of sentences. In their approach, RDF graphs are exploited to represent classes and the relationship of classes in the source code.
Their generated summaries mainly describe what is the code concern and how it is implemented.%leveraging the above information the sentences are produced, which is comprising the summary comments}

%{In 2011, Rastkar et al.\cite{rastkar2011generating} exploit RDF model to represent the structural information for programs code, finally generate the natural language summary for the crosscutting source code concerns.}

\subsubsection{Stereotype Identification based Comment Generation Algorithms}

Stereotypes are a simple abstraction of a class's role and responsibility in a system's design, for instance, an accessor is one of the method stereotype that returns information\cite{dragan2010automatic}. 

Moreno and Aponte et al. \cite{moreno2013jsummarizer,moreno2013automatic} use stereotype information of Java classes and methods to select content to be included into the summary and generate summary for Java class. First, they identify stereotype information of Java classes and methods. Next they exploit heuristic rules % for selection of contents which will be contained in the summary.   
to determine which information in source code to be extracted and added into the summaries. Their summaries focus on the responsibilities and content of the classes instead of their relationships with other classes. In another effort, Abid et al. \cite{abid2015using} differentiate the type of stereotype of C++ methods using stereotype identification, and employ static analysis to extract the main components of the method, so as to generate summaries according to predesigned templates for every type of stereotype of methods.
 
%Stereotype is a kind of extension mechanism in Unified Modeling Language (UML). Here  it is a
%\begin{figure*}[t!]
%	\centering
%	\includegraphics[width=0.7\textwidth]{figure5.pdf}
%	\caption{The classification of quality evaluation approaches and criteria of code comments.}
%	\label{fig5}
%\end{figure*} 
\subsubsection{Other Algorithms}

Employing heuristics approach only, Sridhara and Pollock et al. \cite{sridhara2011automatically} generate concise comments for the high-level action in Java methods. In their commenting system, they design and implement the rule sets by which code snippets of statement sequences, conditionals and loops are identified. Next they develop the corresponding summary templates for generating comments for Java methods. Finally they generate summaries for Java methods according to these templates. 

Among many literature on the study of commenting generation, there are several studies which adopt approaches initially designed for non-commenting problems. In a different effort, Rodeghero et al. \cite{rodeghero2014improving} introduce an eye-tracking technique into comment generation researches. Basing on the related studies of eye-tracking in program comprehension, they detect eye movements and the amount of gaze time which programmers spend in scanning source code. According to these information they adjust the weight value of words in source code, and identify keywords, which makes generated summaries more accurate. 
In another effort, Wang  and Pollock et al. \cite{wang2017automatically} use some rules mined from open source projects to identify the object-related action unit within a method, and generate natural language phrases for action unit. They exploit data-driven approach to obtain a set of rules to identify the focal statements and arguments in source code. In the end, they employ the information obtained in the previous step to construct natural language description for action unit according to the predefined templates.% And some rules can be used to improve natural language phrase.}

%{Fowkes et al. \cite{fowkes2017autofolding} propose an autofolding method for source code summarization based on optimizing the similarity between the summary and the source file. Their method automatically creates a code summary by folding less informative code regions. They formulate code summary as a sequence of AST folding decision, leveraging a scoped topic model for code tokens.}
\iffalse
The above are the taxonomies of the existing comment generation systems and algorithms at present. In the following we further summarize the work of features of various algorithms.%deep neural networks are one sort of commenting technologies.

Most of aforementioned algorithms require high quality datasets. At present, the different commenting systems are equipped with different training datasets. High quality datasets are qualified at the big scale and the diversity in terms of programming language and various domains. that is the quantity of dataset is big enough to meet all the data needs of commmenting system. Most of existing comment generation systems select the open source project from three main data sources: some select data from software repository in GitHub which contain code segments and corresponding comments, some select the posts from Q\&A site in Stack Overflow, which is used to build dataset containing code segments and some textual discussion or comments. Others select data from developers' communications, such as bug reports and emails. 
\fi
\subsection{Summarization OF COMMENT GENERATION ALGORITHMS}%\shl{Currently, this section does not really compare the algorithms. You just describe the characteristics of those algorithms. I think this section can be removed if you want to submit it soon.}
We have described three main categories of comment generation algorithms, and we further summarize the characteristics of those algorithms.  %Which one is the best in performance? At present, there is no study to compare these algorithms exclusively. The difficulty of the comparison task lies in that the lack of unifying datasets and the programming language dependency of each commenting algorithm. However, each type of algorithm is characterized by different features. %Two studies are performed to make a comparison of the algorithms on the other software engineering tasks, such as the code-suggestion tasks \cite{hellendoorn2017deep}
The characteristics of Information Retrieval (IR) based algorithms are salient on three aspects. First, these kinds of commenting algorithms view source code as plain text, and construct comments by searching keywords or tokens from source code or searching comments from similar code. They attach great importance to lexical semantics of source code, and neglect the structural information, data dependency and invoke information of source code. Second, the effectiveness of IR-based algorithms, to a large extent, depends on the similar code from datasets. Without similar code in datasets, IR-based algorithms will fail to generate accurate comments or any comment at all for classes. If the source code contains poorly named identifiers, they still fail to recommend accurate comments. Third, for IR-based algorithms, the quality and quantity of generated comments both depend on the quality and quantity of source code contained in datasets.  %find out the similar code by the way of VSM , LSI, code clone techniques and the probability model, such as LDA etc. These techniques can generate comments for target code 

Among deep neural network based comment generation algorithms, most of them employ RNN encoder-decoder model with the assistance of attention mechanism to represent the features of input code as the vector. They usually formulate comments generation from source code as machine translation in natural language processing. The role of attention mechanism lies in adjusting weights of the tokens in source code, which makes the accuracy of the generated comments better. When choosing CNN to build comment generation system, it is necessary to equip it with attention mechanism. The deep neural network based comment generation algorithms are one type of the supervised learning. So, these kinds of algorithms require high quality datasets for training parameters of neural network so as to acquire the generative model for source code. % or add CNN structure to encoder-decoder model to generate comment for class, method and code block etc, or simply CNN structure equipped with attention model to represent the structure and lemix of source code, . 

Besides information retrieval based algorithms and deep neural network based algorithms, other comment generation algorithms mostly either adopt some existing models from other research areas to represent source code, such as eye-tracking and semantic ontology based RDF etc., or adopt models that exclude the aforementioned models, such as SWUM model to represent source code and yield satisfactory comments for given source code.

%\shl{It is weird to simply mention these 5 papers. You should try to summarize the algorithms presented in Section III.D. You donot need to cite new papers here because they are already cited before.} 

%Generating high quality comments is the goal for researchers who engage in the researches on comment generation, the quality of comments is the important metric to evaluate the performance of commenting systems and comment generation algorithms. Accordingly, the studies on evaluation of comment quality are closely correlate to the researches on the automatic comment generation.

%% file: quality.tex
\section{QUALITY EVALUATION CRITERIA OF CODE COMMENTS}
\label{sec:quality}
As discussed above, most existing commenting algorithms are evaluated based on different datasets, which causes the experimental results to be noncomparable. The fundamental reasons lie in the lack of an unified dataset for algorithm evaluation. In addition, appropriate comment evaluation criteria in terms of the efficiency and effectiveness of algorithms are still missing.% Whether automatic comment generation algorithm is effective and efficient requires objective and scientific evaluation criteria and unified validation dataset to test.

\begin{figure*}[t!]
	\centering
	\includegraphics[width=0.85\textwidth]{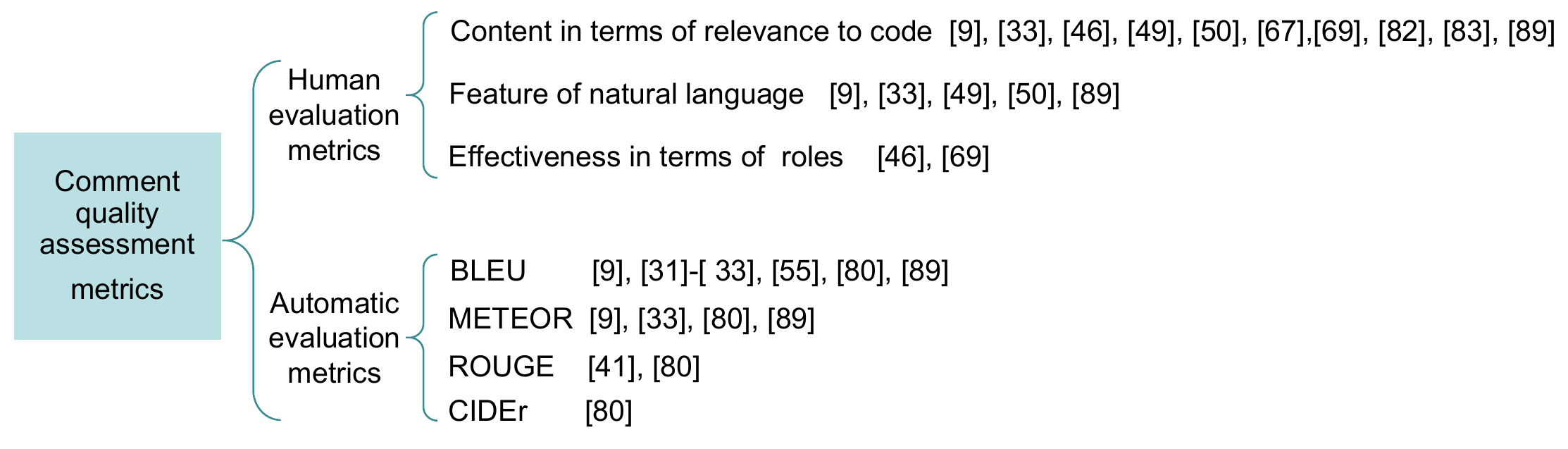}
	\caption{The classification of code comment quality assessment metrics and referenced literature.}
	\label{fig5}
\end{figure*}

In recent years, some researchers have conducted many studies on comment evaluation models. The studies on comment quality assessment and reference literature are shown in Figure 5. Khamis et al. \cite{khamis2010automatic} present JavadocMiner, a tool that can analyze the quality of Javadoc comments automatically. Using a set of heuristics, they evaluate the quality of comments in terms of quality of language and consistency between source code and its comments. %But they simple take the structural requirements of Javadoc into consideration to detect the inconsistencies between code and comments. Apparently this is not enough.
Daniela Steidl et al. \cite{steidl2013quality} perform the study on quality evaluation for native comments with a quality assessment model and a set of comprehensive and systematic metrics. They classify comments into seven categories: copyright comments, header comments, member comments, inline comments, section comments, code comments and task comments. Then they propose the corresponding evaluation metrics and a general quality model for each type of comments according to the code-comment relevance and the length of comments. Their work mainly focuses on the quality of inline comments and member comments. In 2016, Yu et al. \cite{b5} propose a source code quality assessment approach based on aggregation of classification algorithms, which evaluates comments in terms of their format, language form, contents and correlation degree of code. This method is characterized by introducing machine learning and natural language processing technologies into comment quality assessment. They improve the evaluation results by using the aggregation of the basic classification algorithms. In general, there are a few studies on quality assessment for source code comments in existing literature, however, there is a lack of studies on the quality assessment of analytical comments. % quality assessment are much less, so the corresponding results are little.}

There are two classes of  assessment methods available: manual evaluation and automatic evaluation. %(see Figure 5). 
Manual evaluation relies on experienced developers to analyze and rate the generated comments one by one according to predefined quality metrics, such as conciseness, readability, accuracy, etc. Although the manual assessment methods avoid the complex technological design of feature selection and evaluation algorithms, it is usually time-consuming and costly, thus it is not fit for a vast number of analytical comments. However, there are no automatic quality assessment tool available exclusively for code comments, and researchers in software engineering field generally borrow natural language assessment criteria and tools. The commonly used metrics and tools in automatic assessment are BLEU \cite{papineni2002bleu:} and METEOR \cite{banerjee2005meteor:}. Sometimes, ROUGE and CIDER are also used. To sum up, the comment quality assessment generally involves the following three perspectives:% to evaluate quality for generated comments from some commenting systems.} Only with a scientific comment quality assessment model can evaluate commenting systems objectively, which can advance development of automatic comment generation algorithms. On the other hand, we can well know the pros, cons and restrictions of existing code comment generation algorithms and systems, so that facilitating to select suitable algorithms for particular application.

\begin{itemize}
    \item \textbf{Experimental datasets}, which refers to datasets that are used for evaluating code commenting algorithms. %Dataset used in practice is usually called corpus.
    \item  \textbf{Evaluation methods}, which generally include manual assessment and automatic assessment.
    \item  \textbf{Evaluation metrics}, which refer to the criteria for evaluating the quality of comments.
\end{itemize}

\subsection{DATASETS FOR VALIDATION}
The datasets for validation in existing studies usually varies across commenting systems. These datasets come from three sources: open source projects from GitHub software repositories, Q\&A sites in Stack Overflow, and communication and discussion information, like emails, among developers.

Although there only exist three kinds of datasets, every commenting system collects different projects written in different programming languages as datasets. Even some systems collect information from the same source such as Q\&A sites in Stack Overflow, there still exist differences in tags and time segments. As a whole, there is not a uniform public validation dataset for use.

\subsection{AUTOMATIC EVALUATION}
Since comments are sentences written in natural languages, the comment quality assessment criteria mostly come from the corresponding evaluation metrics in the field of natural language processing. These metrics originally are designed for measuring the quality of sentences generated by machine translation system, and have been justified to well reflect the accuracy of test results, and the test results are highly coherent with human evaluation. Existing automatic evaluation metrics for comment quality mainly include BLEU, METEOR, ROUGE and CIDEr.

\subsubsection{BLEU}
BLEU \cite{papineni2002bleu:}, standing for Bilingual Evaluation Understudy, is a method for automatic evaluation of machine translation, proposed by Papineni et al. in 2002. It can be used to analyze automatically the degree of common appearance of candidate texts and reference translations. Since human evaluations of machine translation are extensive and expensive, BLEU is designed as a quick, inexpensive and language-independent automatic evaluation method for machine translation. In terms of the reliability of evaluation results, the results of BLEU highly cohere with human evaluation, so it is extensively applied to the evaluation of machine translation and the evaluation understudy of automatic commenting systems\cite{hu2018deep,hu2018summarizing,zheng2017code}. The key assessment contents of BLEU is n-gram precision, which refers to the proportion of the matched n-grams out of the total number of n-grams in the evaluated translation. Precision is calculated separately for each n-gram order, and the final precision is computed as weighted geometric mean for each precision. %The central design intent of n-gram precision is that 
For the candidate texts, the more similar they are to the natural language reference written by human, the higher scores they get. For n-gram precision, n can be 1, 2, 3, 4. When n=1, BLEU has good performance at the sentences with well matched in corpus level, but the matches in sentence-level become poorer as the length of the sentences grow longer. BLEU does not take direct recall into account. That lack of recall and no consideration about explicit word-matches between translation and reference makes the degree of coherence between human evaluation and BLEU evaluation advance difficultly.   
\subsubsection{METEOR}   
METEOR \cite{banerjee2005meteor:}, short for Metric for Evaluation of Translation with Explicit Ordering, is proposed by Lavie in 2005. It is a  supplementary evaluation metric to BLEU, which incorporates the recall to reflect how much the translated results cover the entire contents of the source sentences. In consideration of the weakness of BLEU method without reflecting the recall, METEOR takes the recall into account, and computes a score based on explicit word-to-word matches between translation and a reference translation and evaluates the translation according to the score. The higher the score, the closer it is to the reference translation, which denotes the higher quality of the translation. Experiments show \cite{banerjee2005meteor:} that METEOR has significantly improved coherence between human judgments in comparison of BLEU. So, METEOR metrics usually are used to evaluate the quality of machine-generated comments of code as the complement to BLEU.

\subsubsection{ROUGE}  
ROUGE (Recall-Oriented Understudy for Gisting Evaluation) proposed by Chin-Yew Lin \cite{lin2004rouge:}, is an automatic summary assessment method, and now it is extensively adopted in the tasks of summary assessment. ROUGE includes several different metrics, among which four commonly used ones are ROUGE-N, ROUGE-L, ROUGE-W and ROUGE-S. They are recall-based methods which measure similarity between a summary and the ideal summaries written by human. %Similar to BLEU, ROUGE has no Fmean\footnote{Fmean: a weighted combination of Precision and Recall.} evaluation metrics, it mainly examines the adequacy and fidelity of system translations, and cannot evaluate the fluency of translation. It computes the n-grams co-occurrence frequency. 
ROUGE-N and ROUGE-L are generally used as the metrics for measuring the summary comment quality assessment\cite{liang2018automatic,wan2018improving}. ROUGE-L denotes the longest common subsequence-based precision and recall statistics.
\subsubsection{CIDEr}  
CIDEr \cite{vedantam2015cider:}, standing for Consensus-based Image Description Evaluation, is proposed by Vedantm in 2015 in the computational vision and mode recognition conference. Researchers believe that multiple existing evaluation metrics have high relevance to human evaluation,  but there is no way to unify them into an uniform metric which can determine the similarity between a candidate sentence and reference sentences. In order to solve this issue, %The aforementioned metrics, BLEU, METEOR and ROUGE, come from the machine translation in natural language processing. Different from these evaluation metrics, CIDER comes from image processing. It is a metric which measures the quality of image description for a given image. It is common for a image description to be assessed that the multiple evaluation results corresponding to different metrics are reported. But there is no uniformed metric for the final evaluation result. %R
a consensus-based metric, named CIDEr, is proposed. Its primary principle is to measure the similarity between a test sentence and the majority of reference sentences. The experiment results demonstrate that CIDERr metric coheres with human consensus-based matches better than the existing aforementioned metrics.

Accordingly, researchers generally exploit BLEU to evaluate machine-produced translations automatically firstly, then METEOR is used as a supplementary to final results of evaluation, which makes assessment results more objective and closer to human evaluation results. And some code summary generation systems adopt ROUGE and CIDEr metrics to evaluate generated comments \cite{wan2018improving}, but ROUGE and CIDEr are not used extensively.

In summary, because of similarities between code commenting and machine translation, researchers usually adopt BLEU, METEOR, ROUGE and CIDEr methods to measure the quality of computer-produced comments. Among these metrics BLEU and METEOR are the common used metrics.

\subsection{HUMAN EVALUATION}
In the commenting systems summarized in Section \ref{sec:algorithm}, researchers ask human evaluators to evaluate the generated comments on assessment metrics. %Fewer systems evaluate the generated comments automatically, besides human evaluation, which mainly happen in deep neural network based commenting systems.

%Human evaluation refers to that the skilled programmers acting as evaluators to perform the comparison between comments generated by systems and references, and calculate scores according to predesigned questions and content metrics, ultimately get assessment results, developers assess the pros and cons of particular algorithm according to the assessment results. 
Though the scores from human evaluation are subjective and manual scoring is subject to low efficiency, it is still one of the important methods for assessing the performance of many commenting algorithms. There are several manual assessment metrics for measuring the quality of generated comments. Although the names of the quality assessment metrics in different researches are not completely the same, we group them into three categories according to their characteristics as follows.

First, measuring code comments with their contents: evaluating the generated comments on contents, such as adequacy, accuracy \cite{sridhara2010towards,mcburney2014automatic,sridhara2011generating}, conciseness, informativeness \cite{iyer2016summarizing} and interpretability \cite{zheng2017code}, etc. The meanings of the features in this group are described as follows.
\begin{itemize}
    \item  \textbf{Accuracy} measures to what extent the generated comments reflect the semantics of the relevant source code.%: refers to evaluate comment text by judging to what extent the generated comment to represent the code‘s actions.
    \item  \textbf{Content adequacy} is used to evaluate how much information the comments miss with regard to the information contained in source code.%:refers to evaluate comment by comparing the comment in content with source code to judge to what extent the comment miss information contained in code.
    \item  \textbf{Conciseness} is used to evaluate to what extent the comments contains the unnecessary information.
    \item  \textbf{Informativeness} measures the amount of contents carried over from the input code to the NL comment, ignoring fluency in language. 
    \item  \textbf{Interpretability} measures to what extent the generated comments convey the meaning of the source code.
\end{itemize}    
 
Second, measuring code comments with the features of natural languages: evaluate the generated comments on grammaticality and fluency, ignoring contents of comments. These modalities are, namely, expressiveness \cite{moreno2013jsummarizer}, naturalness \cite{iyer2016summarizing} and understandability \cite{zheng2017code} etc. The definition of each feature in this group is described as follows.
\begin{itemize}
    \item  \textbf{Naturalness} measures the grammaticality and fluency of comments.
    \item  \textbf{Expressiveness} measures the comments' readability and understandability in the respect of their way of description.
    \item  \textbf{Understandability} is used to evaluate comments according to their fluency and grammar.
\end{itemize}
Naturalness and understandability metrics almost have the same meaning in different names in different researches.

Third, measuring code comments in terms of their effectiveness: judging whether the generated comments are useful \cite{mcburney2014automatic} and necessary \cite{sridhara2011generating}, or evaluating to what extent can the generated comments help developers understand the programs, namely code understandability. The meanings of features in this group are described as follows.
\begin{itemize}
    \item  \textbf{Usefulness} measures to what extent the generated comments are useful for developers to understand code.
    \item  \textbf{Code understandability} is used to evaluate to what degree does the generated comments help developers understand the programs.
    \item  \textbf{Necessity} measures to what extent the generated comments are necessary.
    \item  \textbf{Utility} measures to what extent the generated comment can help developers comprehend code.
\end{itemize}

All modalities discussed above can be rated in different scale. Some researchers rate comments on a scale between 1 and 5; some rate comments on a scale between 1 and 4, or on a scale between 1 and 3. For accuracy metric, evaluators can rate comments by answering multiple choice questions, and each question could be answered as ``Strongly Agree'', ``Agree'', ``Disagree'', ``Strongly Disagree'', four items in total; or answered as ``Strongly Agree'', ``Agree'', ``Neural'', ``Disagree'', ``Strongly Disagree'', five items in total; or ``Accurate'', ``Slightly Inaccurate'', ``Very Inaccurate'', three items in total. The answers are assigned on a scale between 1 and 5, or between 1 and 4, or between 1 and 3. For other metrics, the same approach can be adopted to rate comments.

In addition, there are two assessment metrics worth mentioning: precision and recall. These two metrics can be used to evaluate the usefulness of generated comments. Precision and recall metrics both come from information retrieval metrics \cite{nazar2016summarizing}. Precision \cite{banerjee2005meteor:} is the proportion of the correct n-grams in the evaluated comments. %Here, \sxt{ the matched n-grams refer to n-grams in evaluated comment matched with the reference translation which is generally written by human.} 
Recall \cite{banerjee2005meteor:} is the proportion of the correctly predicted n-grams in reference comments. Among aforementioned metrics, BLEU is based on precision metric, and METEOR is based on recall metric.
 
Evaluators are usually the programmers with five-year experiences in the corresponding programming language development or Masters/PhDs in the corresponding field.

Human evaluation is characterized by high accuracy and convincing assessment results, while it is also subject to high costs, subjective influence from evaluators and being time-consuming. 
However, human evaluation is still a very important quality assessment methods for code comments. And it is not replaced by automatic evaluation. All in all, automatic evaluation has its own advantages, it can supplement the weakness of manual evaluation. Till now, there is no mature, efficient and inexpensive comment quality assessment tools, which is an important problem to solve in the field of code comment generation fields.

To sum up, when designing evaluation experiments and selecting evaluation criteria, we should design and select the suitable assessment criteria according to the algorithms that are adopted in commenting systems so as to make the designed experiments and assessment results more convincing.

%% file: future.tex
\section{FUTURE DIRECTIONS}
\label{sec:future}
As an important research direction in software engineering field, the source code commenting technologies have received much attention from the academics since the last decade. But it remains a challenge research topic due to its internal complexity and the limitations of existing technologies. In recent years, one of the representative efforts is to utilize %was not well solved due to the restrictions in the technologies development. Accordingly, when a new technique is proposed, researchers will try to apply it to the code commenting problem actively, with hope to solve this problem. Recent years, researchers perform a large amount of  work in this area, they exploit 
deep neural networks to solve the problem of automatic comment generation, and some promising results have been obtained. However, there still exist some drawbacks such as low accuracy of generated comments and insufficient generated comments on the whole. In the following discussion, we summarize the future research opportunities for automatic comment generation.%At present, there exist four kinds of potential developments, we will depict them as follows, and hope to provide references for software engineering researchers.
\begin{itemize}
	\item {\textbf{Exploring the synergy between deep neural network and other models.} At present, the deep neural networks technology as the emerging technology is adopted to solve the problem of automatic generation of code comments, and obtains better results than previous methods. %Accordingly, deep neural network is a good approach to automatic comment generation researches. 
The reason is that the structure of deep neural network technology is fit for solving the problem of sequences to sequences. However, the problem of automatic generation of code comments is not simple translation problems within the natural language, it is the conversion problem from structured source code to natural language sentences. As a result, exploring the synergy between deep neural network and other models to represent the source code remains an open research topic.}
\item {\textbf{Fusion of different source code models.} The combination of multiple models for representing source code is well suitable for the solution of code comment generation problems. Because one model is fitter for describing one feature of source code. Token-based model can be used to describe lexical information in source code, which is the words and tokens hidden in identifiers name or comments of source code, such as BoW model etc. And statistical language models describe the probabilities for the words to appear in sequences, such as n-gram model etc. Besides, SWUM model is suitable for representing the semantic, structural information and phrases information in source code, but it cannot represent word sequences in source code. Accordingly, appropriate fusion of different source code models may be open for future research.}
	
\item{\textbf{Designing a customized, intelligent automatic comment generation system to meet various scenarios.} To solve the problem of automatic code comment generation, we should perform mainly two tasks: source code representation and text generation. The goal of code commenting is to improve readability of source code, which is to help software developers and maintenance engineers comprehend programs much faster and better, and is beneficial for them to perform other tasks; On the other hand, code commenting can free developers from writing comments manually, which is laborious and tedious for developers. So it is one of the important directions for automatic code comment generation to design a customized, intelligent automatic comment generation system to meet the requirements of different developers in specific application scenarios.}
\item{\textbf{Unification of test datasets and comment quality assessment model.} In terms of code comment assessment, the unification of multiple test datasets opens the future direction first, because if there were no unified test dataset, there would be no means to compare the advantages and disadvantages of commenting generation algorithms, which will certainly affect the development of algorithms. So designing and building up a unified, universal test dataset is an urgent problem needed to be solved. Second, designing and setting up an appropriate, objective comment quality assessment model is another vital research direction. Accordingly, the study on comment quality assessment metrics is also an active research direction of comment assessment.}
\end{itemize}

%% file: conclusion.tex
\section{CONCLUSION}
\label{sec:conclusion}
This paper provides a  survey of the recent development of automatic code comment generation technology. The research in this field has explored information retrieval based code commenting techniques, neural network based code commenting techniques and other code comment generation techniques. We introduce the features of code commenting problem, the research framework and the workflow of automatic code comment generation. By summarizing and analyzing three main classes of automatic code comment generation techniques, we present the future research opportunities. Specifically, exploring the neural network, combination of multiple technologies and flexible switching in multi-application scenarios remain open research topics.

In the field of quality assessment research on code comments, we summarize four kinds of automatic assessment metrics including BLEU, METEOR, ROUGE and CIDER,  which represent the strengths and weaknesses of each comment metric; and we outline the commonly used metrics in manual evaluation from three aspects: natural language features, contents of comments and effectiveness of comments. Choosing reasonable automatic assessment criteria and manual assessment criteria, building the universal validation datasets are open for future research on comment quality assessment. %We hope the work done in this paper provide a reference for the researches of automatic code comment generation.

%\section{Acknowledgment}
% I would like to thank associate professor Sun Hailong and Dr. Wang Xu for their sustained guidance and support throughout writing process. 